\documentclass[english,english,A4,superscriptaddress, nofootinbib, prd]{revtex4}
\usepackage[T1]{fontenc}
\usepackage[latin9]{inputenc}
\usepackage{color}
\usepackage{babel}
\usepackage{verbatim}
\usepackage{amsbsy}
\usepackage{amstext}
\usepackage{amssymb}
\usepackage[unicode=true,pdfusetitle,
 bookmarks=true,bookmarksnumbered=true,bookmarksopen=true,bookmarksopenlevel=1,
 breaklinks=false,pdfborder={0 0 1},backref=false,colorlinks=true]
 {hyperref}
\hypersetup{
 linkcolor=blue,citecolor=cyan}

\makeatletter

\providecommand{\tabularnewline}{\\}

\@ifundefined{textcolor}{}
{%
 \definecolor{BLACK}{gray}{0}
 \definecolor{WHITE}{gray}{1}
 \definecolor{RED}{rgb}{1,0,0}
 \definecolor{GREEN}{rgb}{0,1,0}
 \definecolor{BLUE}{rgb}{0,0,1}
 \definecolor{CYAN}{cmyk}{1,0,0,0}
 \definecolor{MAGENTA}{cmyk}{0,1,0,0}
 \definecolor{YELLOW}{cmyk}{0,0,1,0}
}

\usepackage{amsmath}

\makeatother

\begin{document}
\title{Spin alignment of vector mesons in local equilibrium by Zubarev's
approach}
\author{Shi-Zheng Yang}
\affiliation{Department of Modern Physics and Anhui Center for fundamental Sciences
(Theoretical Physics), University of Science and Technology of China,
Anhui 230026}
\author{Xin-Qing Xie}
\affiliation{Department of Modern Physics and Anhui Center for fundamental Sciences
(Theoretical Physics), University of Science and Technology of China,
Anhui 230026}
\author{Shi Pu}
\affiliation{Department of Modern Physics and Anhui Center for fundamental Sciences
(Theoretical Physics), University of Science and Technology of China,
Anhui 230026}
\affiliation{Southern Center for Nuclear-Science Theory (SCNT), Institute of Modern
Physics, Chinese Academy of Sciences, Huizhou 516000, Guangdong, China}
\affiliation{Shanghai Research Center for Theoretical Nuclear Physics, NSFC and
Fudan University, Shanghai 200438, China}
\author{Jian-Hua Gao}
\affiliation{Shandong Key Laboratory of Optical Astronomy and Solar-Terrestrial
Environment, School of Space Science and Physics, Shandong University,
Weihai, Shandong, 264209, China}
\author{Qun Wang}
\affiliation{Department of Modern Physics and Anhui Center for fundamental Sciences
(Theoretical Physics), University of Science and Technology of China,
Anhui 230026}
\affiliation{School of Mechanics and Physics, Anhui University of Science and Technology,
Huainan,Anhui 232001, China}
\begin{abstract}
We compute the $00$ element of the spin density matrix, denoted as
$\rho_{00}$ and called the spin alignment, up to the second order
of the gradient expansion in local equilibrium by Zubarev's approach.
In the first order, we obtain $\rho_{00}=1/3$, meaning that the contributions
from thermal vorticity and shear stress tensor are vanishing. The
non-vanishing contributions to $\rho_{00}-1/3$ appear in the second
order of gradients in the Belinfante and canonical cases. We also
discuss the properties of {\normalsize the spin density matrix }under
the time reversal transformation. The effective transport coefficient
for the spin alignment induced by the thermal shear stress tensor
is T-odd in the first order, implying that the first order effect is
dissipative. 
\end{abstract}
\maketitle

\section{introduction}

In non-central heavy-ion collisions there is a huge orbital angular
momentum (OAM) generated in the initial state of two colliding nuclei
at relativistic energies. A small fraction of the initial OAM can
be transferred into the medium in the form of the spin polarization
of quarks (and antiquarks) via spin-orbit couplings, which leads to
the global spin polarization of hadrons with respect to the reaction
plane \citep{Liang:2004ph,Voloshin:2004ha,Liang:2004xn,Gao:2007bc}.
Later, Ref. \citep{Becattini:2007sr} has established a connection
between global spin polarization and hydrodynamic vorticity. 

The global spin polarization of $\Lambda$ hyperons (including $\overline{\Lambda}$)
has been measured by experiments of STAR \citep{STAR:2008lcm,STAR:2017ckg,STAR:2019erd,STAR:2020xbm,STAR:2022fan},
ALICE \citep{ALICE:2023jad,ALICE:2019aid} and HADES \citep{HADES:2022enx}.
These measurements have been successfully described by a variety of
phenomenological models \citep{Becattini:2007sr,Karpenko:2016jyx,Xie:2017upb,Li:2017slc,Sun:2017xhx,Shi:2017wpk,Wei:2018zfb,Xia:2018tes,Vitiuk:2019rfv,Fu:2020oxj,Ryu:2021lnx,Lei:2021mvp,Wu:2022mkr}.
In contrast, the spin polarization along the beam direction as functions
of azimuthal angles \citep{Niida:2018hfw,STAR:2019erd} cannot be
described by the polarization from the thermal vorticity in hydrodynamic
or transport simulations using the Cooper-Frye formula \citep{Becattini:2013fla,Fang:2016vpj}.
Several models have been proposed to explain the spin polarization
along the beam direction, such as spin polarization from the temperature
vorticity \citep{Wu:2019eyi}, the shear-induced polarization \citep{Liu:2021uhn,Liu:2021nyg,Fu:2021pok,Becattini:2021suc,Becattini:2021iol},
spin Hall effects \citep{Liu:2020dxg,Fu:2022myl,Wu:2022mkr}, weak
magnetic fields induced polarization \citep{Sun:2024isb} and corrections
from interactions between quarks and background fields \citep{Fang:2023bbw}.
Very recently, the hydrodynamical simulations with the assumption
of the isothermal equilibrium \citep{Palermo:2024tza} and the results
from blast wave model \citep{Arslan:2024dwi} can give a quantitatively
description to local spin polarization at very high energy collisions.
While, the recent hydrodynamic studies on the spin polarization along
the beam direction at p+Pb collisions disagrees with the data measured
by CMS experiments \citep{Yi:2024kwu}. Although the azimuthal angle
dependence of the spin polarization along the beam direction can be
qualitatively understood by incorporating these effects, its quantitative
description remains a challenge in general \citep{STAR:2019erd,ALICE:2021pzu,STAR:2023eck}.
We refer readers to recent reviews about spin polarization in heavy
ion collisions \citep{Wang:2017jpl,Florkowski:2018fap,Liang:2019clf,Becattini:2020ngo,Huang:2020dtn,Gao:2020lxh,Hidaka:2022dmn,Becattini:2024uha,Niida:2024ntm}.

Meanwhile, the spin alignment of vector mesons in heavy ion collisions
was also proposed \citep{Liang:2004xn}. The spin alignment of vector
mesons is described by the spin density matrix $\rho_{ss^{\prime}}$,
where $s,s^{\prime}=0,\pm1$ represent spin states with the specified
spin quantization direction. The component $\rho_{00}$ can be measured
through the strong decay of vector mesons \citep{Faccioli:2010kd}.
The condition $\rho_{00}>1/3$ or $\rho_{00}<1/3$ corresponds to
longitudinal and transverse polarization (referred to the polarization
vector of the vector meson field) with respect to the spin quantization
direction, respectively. The first measurement of the the global spin
alignment of $\phi$ mesons in Au+Au collisions at $\sqrt{s_{NN}}=11.5-200\ \mathrm{GeV}$
was performed by the STAR collaboration \citep{STAR:2022fan}. The
magnitude of $\rho_{00}-1/3$ for $\phi$ mesons at low energies is
significantly larger than expected. In the same experiment, In contrast,
the global spin alignment of $K^{*0}$ mesons was also measured in
the same experiment but it was found to be consistent with zero within
errors. Recently, the ALICE collaboration also observed the spin alignment
of $J/\Psi$ in Pb+Pb collisions at $\sqrt{s_{NN}}=5.02\ \mathrm{TeV}$
\citep{ALICE:2023jad}. 

The spin alignment of $\phi$ mesons can be described by the relativistic
quark coalescence model incorporated into the spin Boltzmann (kinetic)
equation for vector mesons, where the quark and antiquark are polarized
by the effective $\phi$ field \citep{Sheng:2019kmk,Sheng:2020ghv,Sheng:2022wsy,Sheng:2023urn}.
There are other studies such as the models based on glasma \citep{Kumar:2022ylt,Kumar:2023ghs},
self-energy effects \citep{Fang:2023bbw}, and other second-order
hydrodynamics effects \citep{Kumar:2023ojl}, as well as those employing
the light-cone formalism \citep{Fu:2023qht}. It has been found that
unlike $\phi$ mesons, rescatterings in the medium lead to a decay
in the spin alignment of $\rho$ mesons \citep{Yin:2024dnu}. The
spin alignment of $J/\Psi$ has also been explored within the holographic
model \citep{Zhao:2024ipr,Sheng:2024kgg}. The spin alignment of vector
mesons can also come from the shear viscous tensor \citep{Li:2022vmb,Wagner:2023cct}.
A study of the shear contribution in the quark-gluon plasma \citep{Dong:2023cng}
suggests that such a contribution is not important as compared with
to that from the strong force field. 

In this work, we investigate the spin alignment of vector mesons using
Zubarev's approach \citep{1979TMP....40..821Z,zubarev1981modern,Becattini:2019dxo,Becattini:2015ska}
in quantum statistical field theory. In this approach, hydrodynamical
quantities can be computed by the quantum statistical method in global
and local equilibrium. The approach has alrealy been applied to study
the spin polarization of $\Lambda$ hyperons in both global \citep{Becattini:2013fla,Becattini:2016gvu,Palermo:2021hlf,Palermo:2023cup}
and local equilibrium \citep{Becattini:2021iol,Becattini:2021suc,Buzzegoli:2021wlg}.

The structure of the paper is organized as follows. In Sec. \ref{sec:vector-field-and-density-operator},
we introduce the spin density matrix of vector mesons. In Sec. \ref{sec:Spin-density-matrix_expansion},
we expand the spin density matrix in the gradient expansion. We derive
the $\rho^{00}$ in the first and second order in the gradient expansion
in Sec. \ref{sec:rho00}. We study the spin alignment induced by the
shear tensor under time reversal transformation in Sec. \ref{sec:dissipative}
and summarize in Sec. \ref{sec:summary}. Throughout this work, we
choose the metric $g^{\mu\nu}=\mathrm{diag}\{1,-1,-1,-1\}$ and Levi-Civita
tensor $\epsilon^{0123}=1$. For simplicity, for a four vector $A^{\mu}=(A^{0},\mathbf{A})$,
we define $\tilde{A}^{\mu}=(A^{0},-\mathbf{A})$ and $\overline{A}^{\mu}=(0,\mathbf{A})$.

\section{Vector fields and spin density matrix }

\label{sec:vector-field-and-density-operator}We start from the Lagrangian
for charge nuetral spin-1 particles, 
\begin{eqnarray}
\mathcal{L} & = & -\frac{1}{4}F_{\mu\nu}F^{\mu\nu}+\frac{1}{2}m^{2}A\cdot A-A\cdot j,
\end{eqnarray}
where $j^{\mu}$ represents the source term, and
\begin{eqnarray}
F^{\mu\nu} & = & \partial^{\mu}A^{\nu}-\partial^{\nu}A^{\mu},
\end{eqnarray}
is the field strength tensor for the spin-1 particle. Note that the
$A^{\mu}$ field is real for charge neutral particles. The equation
of motion derived from this Lagrangian, known as the Proca equation,
is 
\begin{eqnarray}
(\partial^{2}+m^{2})A^{\mu}-\partial^{\mu}\partial_{\nu}A^{\nu} & = & j^{\mu}.
\end{eqnarray}
Taking another derivative $\partial_{\mu}$ on both sides of the above
equation, we obtain 
\begin{eqnarray}
\partial_{\mu}A^{\mu} & = & \frac{1}{m^{2}}\partial_{\mu}j^{\mu}.
\end{eqnarray}
If the source current is conserved, $\partial\cdot j=0$, the above
equation leads to a constraint equation
\begin{eqnarray}
\partial\cdot A & = & 0.\label{eq:constraint_01}
\end{eqnarray}
Under this condition, the Proca equation can be reduced to the Klein-Gordon
equation for each component of the field,
\begin{eqnarray}
(\partial^{2}+m^{2})A^{\mu} & = & j^{\mu}.\label{eq:EoM_02}
\end{eqnarray}

In free space without the source, $j^{\mu}=0$, the solution to the
Proca equation under the constraint (\ref{eq:constraint_01}) can
be expressed as 
\begin{eqnarray}
A_{\mu}(x) & = & \int\frac{d^{3}p}{(2\pi)^{3}}\frac{1}{\sqrt{2E_{\mathbf{p}}}}\sum_{s=\pm1,0}\left[a_{\mathbf{p}}^{s}\epsilon_{\mu}^{s}(\mathbf{p})e^{-ip\cdot x}+a_{\mathbf{p}}^{s\dagger}\epsilon_{\mu}^{s*}(\mathbf{p})e^{ip\cdot x}\right],\label{eq:vector_field}
\end{eqnarray}
where $s=\pm1,0$ denote three polarization states, and $\epsilon_{\mu}^{s}(\mathbf{p})$
is the polarization four-vector corresponding to the spin state $s$
and the on-shell momentum $p^{\mu}=(E_{\mathbf{p}},\mathbf{p})$ with
$E_{\mathbf{p}}=\sqrt{|\mathbf{p}|^{2}+m^{2}}$, (through it is denoted
as a function of the three-momentum $\mathbf{p}$). Here, the creation
and annihilation operators satisfy the following commutation rule,
\begin{eqnarray}
[a_{{\bf p}}^{r},a_{{\bf q}}^{s\dagger}] & = & (2\pi)^{3}\delta^{3}({\bf p}-{\bf q})\delta_{rs}.\label{eq:commutation-rule}
\end{eqnarray}
Note that the convention in the expansion in (\ref{eq:vector_field})
is different from Eq. (5) of Ref. \citep{Sheng:2022ffb} by a factor
$\sqrt{2E_{\mathbf{p}}}$ so the the commutation rule (\ref{eq:commutation-rule})
is different from Eq. (6) of Ref. \citep{Sheng:2022ffb} by a factor
$2E_{\mathbf{p}}$. The constraint equation (\ref{eq:constraint_01})
becomes, 
\begin{eqnarray}
p^{\mu}\epsilon_{\mu}^{s}({\bf p}) & = & 0,
\end{eqnarray}
which implies that $\epsilon_{\mu}^{s}(\mathbf{p})$ is space-like
and has only three independent components. In the particle's rest
frame, the polarization vector is $\epsilon_{s}^{\mu}=(0,\mathbf{e}_{s})$
or $\epsilon_{\mu}^{s}=(0,-\mathbf{e}_{s})$, where $\mathbf{e}_{s}$
is a unit vector for the spin state $s$. For instance, if the quantization
direction is along the $y$-axis, we have 
\begin{eqnarray}
\mathbf{e}_{0} & = & (0,1,0),\nonumber \\
\mathbf{e}_{+1} & = & -\frac{1}{\sqrt{2}}(i,0,1),\nonumber \\
\mathbf{e}_{-1} & = & -\frac{1}{\sqrt{2}}(i,0,-1).\label{eq:n_vector}
\end{eqnarray}
These polarization vectors in the rest frame can be transformed to
the laboratory frame by the Lorentz transformation,
\begin{eqnarray}
\epsilon_{s}^{\mu}(\mathbf{p}) & = & \left(\frac{\mathbf{p}\cdot\mathbf{e}_{s}}{m},\mathbf{e}_{s}+\frac{\mathbf{p}\cdot\mathbf{e}_{s}}{m(E_{\mathbf{p}}+m)}\mathbf{p}\right).\label{eq:spin_vector}
\end{eqnarray}
One can check that these polarization vectors satisfy orthogonal and
completeness relations 
\begin{eqnarray}
\epsilon_{\mu}^{s}(\mathbf{p})\epsilon_{s^{\prime}}^{\mu*}(\mathbf{p}) & = & -\delta_{ss^{\prime}},\label{eq:spin_orthogonality}\\
\sum_{s}\epsilon_{s}^{\mu}(\mathbf{p})\epsilon_{s}^{\nu*}(\mathbf{p}) & = & -\left(g^{\mu\nu}-\frac{p^{\mu}p^{\nu}}{m^{2}}\right).\label{eq:spin_completeness}
\end{eqnarray}

We can define the Winger function for the vector fields as 
\begin{eqnarray}
W^{\mu\nu}(x,k) & = & \int d^{4}y\left\langle A^{\nu}\left(x-\frac{y}{2}\right)A^{\mu}\left(x+\frac{y}{2}\right)\right\rangle e^{ik\cdot y},
\end{eqnarray}
where $\left\langle \hat{O}\right\rangle \equiv\mathrm{Tr}(\hat{\varrho}\hat{O})$
with $\hat{\varrho}$ being the density operator to be defined shortly.
Inserting Eq. (\ref{eq:vector_field}) for $A^{\mu}$ fields into
the above formula, at the leading order in $\hbar$, we obtain 
\begin{eqnarray}
W^{\mu\nu}(x,k) & = & 2\pi\delta(k^{2}-m^{2})\nonumber \\
 &  & \times\sum_{r,s}\{\theta(k_{0})\epsilon_{s}^{\nu*}(\mathbf{k})\epsilon_{r}^{\mu}(\mathbf{k})f_{sr}(x,\mathbf{k})\nonumber \\
 &  & +\theta(-k_{0})\epsilon_{s}^{\nu}(-\mathbf{k})\epsilon_{r}^{\mu*}(-\mathbf{k})[\delta_{rs}+f_{rs}(x,-\mathbf{k})]\},\label{eq:Wigner_function_01}
\end{eqnarray}
where $\delta_{rs}$ stands for the vacuum contribution, and $f_{sr}(x,\mathbf{k})$
is the $sr$ component of the matrix valued spin dependent distribution
function (MVSD) \citep{DeGroot:1980dk,Becattini:2021lfq,Becattini:2020sww,Sheng:2021kfc},
\begin{eqnarray}
f_{sr}(x,\mathbf{k}) & = & \int\frac{d^{3}\mathbf{q}}{(2\pi)^{3}}e^{-iq\cdot x}\textrm{Tr}\left(\hat{\varrho}a_{\mathbf{k}-\mathbf{q}/2}^{s\dagger}a_{\mathbf{k}+\mathbf{q}/2}^{r}\right),\label{eq:distribution_sr_a}
\end{eqnarray}
where $q^{\mu}=(E_{\mathbf{k}+\mathbf{q}/2}-E_{\mathbf{k}-\mathbf{q}/2},\mathbf{q})$.
Note that the MVSD in (\ref{eq:distribution_sr_a}) is different from
that defined in Eq. (12) of Ref. \citep{Sheng:2022ffb} by a factor
$2E_{\mathbf{k}}$ and a transpose in spin space. The factor $2E_{\mathbf{k}}$
is due to the convention in the expansion (\ref{eq:vector_field})
and the corresponding commutation rule (\ref{eq:commutation-rule}).
The spin density matrix $\rho$ for the vector meson is just the normalized
MVSD, 
\begin{eqnarray}
\rho_{sr}(x,\mathbf{k}) & \equiv & \frac{f_{sr}(x,\mathbf{k})}{\textrm{Tr }f(x,\mathbf{k})}.\label{eq:spin_density_matrix}
\end{eqnarray}
In general, the MVSD can be defined in a variety of ways. We will
present an alternative definition for the MVSD in Sec.~\ref{sec:another=000020distribution=000020function}.

In this work, we adopt density operators in Zubarev's approach, following
Refs.~\citep{1979TMP....40..821Z,Becattini:2015ska,Becattini:2019dxo,Becattini:2018duy}.
The local equilibrium density operators have different forms corresponding
to different pseudo-gauges. For example, the Belinfante or canonical
forms of density operators are 
\begin{eqnarray}
\hat{\varrho}_{B} & = & \frac{1}{Z_{B}}\exp\left[-\int_{\Sigma}d\Sigma_{\mu}\hat{T}_{B}^{\mu\nu}(y)\beta_{\nu}(y)\right],\label{eq:density_operator_B}\\
\hat{\varrho}_{C} & = & \frac{1}{Z_{C}}\exp\left\{ -\int_{\Sigma}d\Sigma_{\mu}\left[\hat{T}_{C}^{\mu\nu}(y)\beta_{\nu}(y)-\frac{1}{2}\Omega_{\lambda\nu}(y)\hat{S}^{\mu\lambda\nu}(y)\right]\right\} ,\label{eq:density_operator_C}
\end{eqnarray}
where $\beta^{\mu}=u^{\mu}/T$ with $u^{\mu}$ being the fluid velocity,
$\Omega_{\lambda\nu}$ is the spin potential, and $\Sigma$ is a hypersurface.
The subscripts ``$B$'' and ``$C$'' represent the Belinfante
and canonical forms of energy-momentum tensors which are related by
pseudo-gauge transformation 
\begin{eqnarray}
\hat{T}_{C}^{\mu\nu} & = & -F^{\mu\alpha}\partial^{\nu}A_{\alpha}+g^{\mu\nu}\left(\frac{1}{4}F_{\alpha\beta}F^{\alpha\beta}-\frac{1}{2}m^{2}A_{\alpha}A^{\alpha}\right),\nonumber \\
\hat{T}_{B}^{\mu\nu} & = & \hat{T}_{C}^{\mu\nu}+\frac{1}{2}\partial_{\lambda}(\hat{S}^{\lambda\mu\nu}-\hat{S}^{\mu\lambda\nu}-\hat{S}^{\nu\lambda\mu}).\label{eq:relation_TB_TC}
\end{eqnarray}
In Eq.~(\ref{eq:density_operator_C}), the rank-$3$ spin tensor
$\hat{S}^{\mu\alpha\beta}$ is given by 
\begin{equation}
\hat{S}^{\mu\alpha\beta}=F^{\mu\beta}A^{\alpha}-F^{\mu\alpha}A^{\beta}.\label{eq:R3_spin_tensor}
\end{equation}
As mentioned, we consider real vector fields with vanishing current
density in Eqs.~(\ref{eq:density_operator_B}, \ref{eq:density_operator_C}).
If we aim to compute the spin density matrix at the point $x$, then
$\Sigma$ is the hypersurface passing $x$ \citep{van1982maximum,Becattini:2021suc}.
The choice of $\Sigma$ depends on what kind of physics we would like
to address, e.g., if we take $\Sigma$ as the freeze-out hypersurface,
$\rho(x,\mathbf{k})$ represents the spin density matrix on that hypersurface.


\section{Spin density matrix for vector mesons in gradient expansion }

\label{sec:Spin-density-matrix_expansion}In the previous section,
we show in Eq.~(\ref{eq:spin_density_matrix}) that the spin density
matrix for the vector meson is actually the normalized MVSD which
can be evaluated by Eq.~(\ref{eq:distribution_sr_a}) involving the
density operator $\hat{\varrho}$. In this section, we will expand
$\hat{\varrho}$ in terms of creation and annihilation operators and
compute the MVSD through Eq.~(\ref{eq:distribution_sr_a}). 


As shown in Eqs.~(\ref{eq:density_operator_B}, \ref{eq:density_operator_C}),
$\hat{\varrho}$ depends on $\beta^{\mu}$ and $\Omega_{\lambda\nu}$.
As a common strategy, we assume that $\beta^{\mu}(y)$ varies slowly
in the region where the correlation length of the system is much larger
than the distance $|y-x|$. So $\beta^{\mu}(y)$ can be expanded in
gradients \citep{Becattini:2021suc}, 
\begin{eqnarray}
\beta_{\mu}(y) & = & \beta_{\mu}(x)+\partial_{\lambda}\beta_{\mu}(x)(y-x)^{\lambda}+\mathcal{O}(\partial^{2}).\label{eq:Taylor_expansion_beta}
\end{eqnarray}
The symmetric and anti-symmetric parts of $\partial_{\mu}\beta_{\nu}$
are the thermal shear and vorticity tensors respectively,
\begin{eqnarray}
\xi^{\mu\nu} & = & \frac{1}{2}(\partial^{\mu}\beta^{\nu}+\partial^{\nu}\beta^{\mu}),\nonumber \\
\omega^{\mu\nu} & = & -\frac{1}{2}(\partial^{\mu}\beta^{\nu}-\partial^{\nu}\beta^{\mu}).
\end{eqnarray}
It has been widely discussed that the spin potential is proportional
to the thermal vorticity tensor in global equilibrium, 
\begin{eqnarray}
\Omega^{\mu\nu} & = & -\frac{1}{2}(\partial^{\mu}\beta^{\nu}-\partial^{\nu}\beta^{\mu})\sim\mathcal{O}(\partial^{1}).\label{eq:Omega_1}
\end{eqnarray}
For further discussions on the spin potential, we refer the readers
to Ref.~\citep{Becattini:2018duy} in quantum statistical field theory
and to Refs.~\citep{Hattori:2019lfp,Fukushima:2020ucl,Wang:2021ngp,Wang:2021wqq,Xie:2023gbo}
in spin hydrodynamics (see, e.g. Ref.~\citep{Becattini:2024uha}
for a recent review). 


In this paper we consider two types of density operators $\hat{\varrho}_{B}$
and $\hat{\varrho}_{C}$ in Eqs. (\ref{eq:density_operator_B},\ref{eq:density_operator_C})
and make gradient expansion of $\hat{\varrho}_{B}$ and $\hat{\varrho}_{C}$
to the linear order. Inserting Eqs. (\ref{eq:Taylor_expansion_beta})
and (\ref{eq:Omega_1}) into Eqs.~(\ref{eq:density_operator_B})
and (\ref{eq:density_operator_C}), we obtain 
\begin{eqnarray}
\hat{\varrho}_{B} & = & \frac{1}{Z_{B}}\exp\left[-\beta_{\nu}(x)\int_{\Sigma}d\Sigma_{\mu}\hat{T}_{B}^{\mu\nu}(y)-\partial_{\lambda}\beta_{\nu}(x)\int_{\Sigma}d\Sigma_{\mu}\hat{T}_{B}^{\mu\nu}(y)(y-x)^{\lambda}+\mathcal{O}(\partial^{2})\right],\label{eq:rho_B_gradient_expation}\\
\hat{\varrho}_{C} & = & \frac{1}{Z_{C}}\exp\left[-\beta_{\nu}(x)\int_{\Sigma}d\Sigma_{\mu}\hat{T}_{C}^{\mu\nu}(y)-\partial_{\lambda}\beta_{\nu}(x)\int_{\Sigma}d\Sigma_{\mu}\hat{T}_{C}^{\mu\nu}(y)(y-x)^{\lambda}\right.\nonumber \\
 &  & \left.+\frac{1}{2}\Omega_{\lambda\nu}(x)\int_{\Sigma}d\Sigma_{\mu}\hat{S}^{\mu\lambda\nu}(y)+\mathcal{O}(\partial^{2})\right].\label{eq:rho_c_gradient_expation-1}
\end{eqnarray}
For convenience, we introduce the momentum operator $\hat{P}^{\nu}$
and an auxiliary operator $\hat{L}^{\mu\nu}$, both related to $\hat{T}_{C}^{\mu\nu}$,
\begin{eqnarray}
\hat{P}^{\nu} & = & \int_{\Sigma}d\Sigma_{\mu}\hat{T}_{C}^{\mu\nu}(y),\label{eq:hat_P}\\
\hat{L}^{\lambda\nu} & = & \int_{\Sigma}d\Sigma_{\mu}(y-x)^{\lambda}\hat{T}_{C}^{\mu\nu}(y).\label{eq:hat_L}
\end{eqnarray}
Replacing $\hat{T}_{B}^{\mu\nu}$ in Eq.~(\ref{eq:rho_B_gradient_expation})
by $\hat{T}_{C}^{\mu\nu}$ in the Eq.~(\ref{eq:relation_TB_TC}),
we can express the density operators as
\begin{eqnarray}
\hat{\varrho}_{B} & = & \frac{1}{Z_{B}}\exp\left[-\beta_{\mu}(x)\hat{P}^{\mu}-\partial_{\lambda}\beta_{\nu}(x)\hat{L}^{\lambda\nu}(x)-\frac{1}{2}\partial_{\lambda}\beta_{\nu}(x)\hat{S}_{B}^{\lambda\nu}+\mathcal{O}(\partial^{2})\right],\label{eq:rho_1}
\end{eqnarray}
and
\begin{eqnarray}
\hat{\varrho}_{C} & = & \frac{1}{Z_{C}}\exp\left[-\beta_{\mu}(x)\hat{P}^{\mu}-\partial_{\lambda}\beta_{\nu}(x)\hat{L}^{\lambda\nu}(x)+\frac{1}{2}\Omega_{\lambda\nu}(x)\hat{S}_{C}^{\lambda\nu}+\mathcal{O}(\partial^{2})\right],\label{eq:rho_2}
\end{eqnarray}
where $S_{B}^{\mu\nu}$ and $S_{C}^{\mu\nu}$ are operators related
to spin tensors 
\begin{eqnarray}
\hat{S}_{B}^{\lambda\nu} & = & \int_{\Sigma}d\Sigma_{\mu}\left[\hat{S}^{\mu\lambda\nu}(y)+\hat{S}^{\nu\lambda\mu}(y)+\hat{S}^{\lambda\nu\mu}(y)\right],\label{eq:hat_S_B}\\
\hat{S}_{C}^{\lambda\nu} & = & \int_{\Sigma}d\Sigma_{\mu}\hat{S}^{\mu\lambda\nu}(y).\label{eq:hat_S_C}
\end{eqnarray}
In Eqs.~(\ref{eq:rho_1}) and (\ref{eq:rho_2}) we neglected second
order terms in gradient expansion (\ref{eq:Taylor_expansion_beta})
and focus on effects related to $\partial_{\mu}\beta_{\nu}$.

Previous studies have shown that the spin polarization vector for
$\Lambda$ and $\bar{\Lambda}$ hyperons in local equilibrium depends
on the choice of the hypersurface $\Sigma$ in Zubarev's approach
\citep{Becattini:2021suc}. Therefore, to study the spin density matrix
in local equilibrium, it is essential to specify the hypersurface
$\Sigma$ in density operators (\ref{eq:density_operator_B}) and
(\ref{eq:density_operator_C}). In this paper, we adopt the isochronous
hypersurface $\Sigma_{0}$ following Refs.~\citep{Becattini:2013vja,Buzzegoli:2017cqy,Becattini:2021suc},
then the integral on $\Sigma_{0}$ is 
\begin{equation}
\int_{\Sigma_{0}}d\Sigma^{\mu}=\int d^{3}yn^{\mu},\label{eq:isochronous=000020hypersurface}
\end{equation}
where $n^{\mu}=(1,0,0,0)$ is the time-directed normal vector for
the $\Sigma_{0}$, so the spin density matrix will depend on $n^{\mu}$. 

Now we deal with the trace in the distribution function (\ref{eq:distribution_sr_a})
\begin{eqnarray}
\left\langle a_{\mathbf{k}-\mathbf{q}/2}^{s\dagger}a_{\mathbf{k}+\mathbf{q}/2}^{r}\right\rangle  & \equiv & \textrm{Tr}\left(\hat{\varrho}_{B,C}a_{\mathbf{k}-\mathbf{q}/2}^{s\dagger}a_{\mathbf{k}+\mathbf{q}/2}^{r}\right)\nonumber \\
 & = & \frac{1}{Z_{B,C}}\textrm{Tr}\left\{ \exp\left[\hat{A}(x)+\hat{B}_{B,C}(x)\right]a_{\mathbf{k}-\mathbf{q}/2}^{s\dagger}a_{\mathbf{k}+\mathbf{q}/2}^{r}\right\} ,\label{eq:av-bc}
\end{eqnarray}
where $Z_{B,C}=\textrm{Tr}\hat{\varrho}_{B,C}$, and 
\begin{eqnarray}
\hat{A}(x) & = & -\beta_{\mu}(x)\hat{P}^{\mu},\label{eq:a-op}\\
\hat{B}_{B}(x) & = & -\partial_{\lambda}\beta_{\nu}(x)\left(\hat{L}^{\lambda\nu}(x)+\frac{1}{2}\hat{S}_{B}^{\lambda\nu}\right),\label{eq:B1_original}\\
\hat{B}_{C}(x) & = & -\partial_{\lambda}\beta_{\nu}(x)\hat{L}^{\lambda\nu}(x)+\frac{1}{2}\Omega_{\lambda\nu}(x)\hat{S}_{C}^{\lambda\nu}.\label{eq:B2_original}
\end{eqnarray}
Hereafter we will suppress indices $B$ and $C$ for simple notations.
Using Eq.~(\ref{eq:commutation-rule}), we obtain following commutators
\begin{eqnarray}
\left[\hat{A},a_{\mathbf{p}}^{s\dagger}\right] & = & -\beta_{\mu}p^{\mu}a_{\mathbf{p}}^{s\dagger},\label{eq:commutation_A+}\\
\left[\hat{A},a_{-\mathbf{p}}^{s}\right] & = & \beta_{\mu}\tilde{p}^{\mu}a_{-\mathbf{p}}^{s},\label{eq:commutation_A-}\\
\left[\hat{B},a_{\mathbf{p}}^{s\dagger}\right] & = & \sum_{r}\Xi_{++}^{sr}(p)a_{\mathbf{p}}^{r\dagger}+\sum_{r}\Xi_{+-}^{sr}(p)a_{-\mathbf{p}}^{r},\label{eq:commutation_B+}\\
\left[\hat{B},a_{-\mathbf{p}}^{s}\right] & = & \sum_{r}\Xi_{-+}^{sr}(p)a_{\mathbf{p}}^{r\dagger}+\sum_{r}\Xi_{--}^{sr}(p)a_{-\mathbf{p}}^{r},\label{eq:commutation_B-}
\end{eqnarray}
where $\hat{B}$ represents $\hat{B}_{B}$ or $\hat{B}_{C}$, and
$\Xi_{ij}^{sr}$ ($i,j=\pm$) are given by Eqs.~(\ref{eq:Xi_Bpp})-(\ref{eq:Xi_Cpm})
for Belinfante and canonical cases with detailed derivations in Appendix~\ref{sec:Detail-of-commutation}.

With the help of Baker-Hausdorff formula 
\begin{eqnarray}
e^{-X}Ye^{X} & = & Y-[X,Y]+\frac{(-1)^{2}}{2!}[X,[X,Y]]\nonumber \\
 &  & +\frac{(-1)^{3}}{3!}[X,[X,[X,Y]]]+\cdots,\label{eq:Baker-Hausdorff_formular}
\end{eqnarray}
and Eqs.~(\ref{eq:commutation_A+} )-(\ref{eq:commutation_B-}),
we derive
\begin{eqnarray}
e^{-\hat{A}-\hat{B}}a_{\mathbf{p}}^{s\dagger}e^{\hat{A}+\hat{B}} & = & \sum_{r}\left(e^{\beta\cdot p}\delta_{sr}+D_{++}^{sr}\right)a_{\mathbf{p}}^{r\dagger}+\sum_{r}D_{+-}^{sr}a_{-\mathbf{p}}^{r}+\mathcal{O}(\partial^{3}),\label{eq:rho_a^dagger_rho}\\
e^{-\hat{A}-\hat{B}}a_{-\mathbf{p}}^{s}e^{\hat{A}+\hat{B}} & = & \sum_{r}\left(e^{-\beta\cdot\tilde{p}}\delta_{sr}+D_{--}^{sr}\right)a_{-\mathbf{p}}^{r}+\sum_{r}D_{-+}^{sr}a_{\mathbf{p}}^{r\dagger}+\mathcal{O}(\partial^{3}),\label{eq:rho_a_rho}
\end{eqnarray}
where $\hat{B}$ represents $\hat{B}_{B}$ or $\hat{B}_{C}$, $\tilde{p}^{\mu}=(E_{\mathbf{p}},-\mathbf{p})$,
and coefficients $D_{ij}^{sr}$, with $i,j=\pm$, are given in Eqs.~(\ref{eq:Dpp})-(\ref{eq:Dmp}).
By applying the cyclic property of trace and using Eqs.~(\ref{eq:rho_a^dagger_rho})
and (\ref{eq:rho_a_rho}), we find

\begin{eqnarray}
\left\langle a_{\mathbf{p}^{\prime}}^{r}a_{\mathbf{p}}^{s\dagger}\right\rangle  & = & \sum_{s^{\prime}}\left(e^{\beta\cdot p}\delta_{ss^{\prime}}+D_{++}^{ss^{\prime}}\right)\left\langle a_{\mathbf{p}}^{s^{\prime}\dagger}a_{\mathbf{\mathbf{p}^{\prime}}}^{r}\right\rangle +\sum_{s^{\prime}}D_{+-}^{ss^{\prime}}\left\langle a_{-\mathbf{p}}^{s^{\prime}}a_{\mathbf{p}^{\prime}}^{r}\right\rangle +\mathcal{O}(\partial^{3}),\label{eq:ensemble_average_relation_1}\\
\left\langle a_{\mathbf{\mathbf{p}^{\prime}}}^{r}a_{-\mathbf{p}}^{s}\right\rangle  & = & \sum_{s^{\prime}}\left(e^{-\beta\cdot\tilde{p}}\delta_{ss^{\prime}}+D_{--}^{ss^{\prime}}\right)\left\langle a_{-\mathbf{p}}^{s^{\prime}}a_{\mathbf{p}^{\prime}}^{r}\right\rangle +\sum_{s^{\prime}}D_{-+}^{ss^{\prime}}\left\langle a_{\mathbf{p}}^{s^{\prime}\dagger}a_{\mathbf{p}^{\prime}}^{r}\right\rangle +\mathcal{O}(\partial^{3}),\label{eq:ensemble_average_relation_2}
\end{eqnarray}
where $\left\langle \cdot\right\rangle $ is defined in (\ref{eq:av-bc}).
Combining Eqs.~(\ref{eq:ensemble_average_relation_1}) and (\ref{eq:ensemble_average_relation_2})
with following relations,
\begin{eqnarray}
\left\langle a_{\mathbf{p}^{\prime}}^{r}a_{\mathbf{p}}^{s\dagger}\right\rangle  & = & \left\langle a_{\mathbf{p}}^{s\dagger}a_{\mathbf{p}^{\prime}}^{r}\right\rangle +(2\pi)^{3}\delta(\mathbf{p}-\mathbf{p}^{\prime})\delta_{rs},\nonumber \\
\left\langle a_{-\mathbf{p}}^{s}a_{\mathbf{p}^{\prime}}^{r}\right\rangle  & = & \left\langle a_{\mathbf{p}^{\prime}}^{r}a_{-\mathbf{p}}^{s}\right\rangle ,\label{eq:commutator}
\end{eqnarray}
we can solve $\left\langle a_{\mathbf{p}}^{s\dagger}a_{\mathbf{p}^{\prime}}^{r}\right\rangle $
as,
\begin{eqnarray}
\left\langle a_{\mathbf{p}}^{s\dagger}a_{\mathbf{\mathbf{p}^{\prime}}}^{r}\right\rangle  & = & (2\pi)^{3}\delta(\mathbf{p}-\mathbf{p}^{\prime})\nonumber \\
 &  & \times\left[(e^{\beta\cdot p}-1)I+D_{++}-D_{+-}\mathcal{A}^{-1}D_{-+}\right]_{sr}^{-1}+\mathcal{O}(\partial^{3}),\label{eq:aa_01}
\end{eqnarray}
where $\mathcal{A}=(e^{-\beta\cdot\tilde{p}}-1)I+D_{--}$ with $I$
being the unity matrix in spin space. Inserting Eq.~(\ref{eq:aa_01})
into Eq.~(\ref{eq:distribution_sr_a}), we derive the expression
for $f_{sr}$ up to $O(\partial^{2})$, 
\begin{eqnarray}
f_{sr}(x,\mathbf{k}) & = & \int d^{3}\mathbf{q}e^{-iq\cdot x}\left[(e^{\beta\cdot(k-q/2)}-1)I+D_{++}-D_{+-}\mathcal{A}^{-1}D_{-+}\right]_{sr}^{-1}\delta(\mathbf{q}),
\end{eqnarray}
where $p^{\mu}=(k-q/2)^{\mu}$. Note that $D_{++}$ and $D_{--}$
contain momentum derivatives.


We then consider the gradient expansion for $f_{sr}(x,\mathbf{k})$,
\begin{eqnarray}
f_{sr}(x,\mathbf{k}) & = & f_{sr}^{(0)}(x,\mathbf{k})+f_{sr}^{(1)}(x,\mathbf{k})+f_{sr}^{(2)}(x,\mathbf{k})+\mathcal{O}(\partial^{3}),\label{eq:gradient_expand_01}
\end{eqnarray}
where the superscripts ``$(i)$'' ($i=0,1,2,\cdots$) denote the
power of the gradient. The zero-th order contribution is 
\begin{eqnarray}
f_{sr}^{(0)}(x,\mathbf{k}) & = & \frac{1}{e^{\beta\cdot k}-1}\delta_{sr}=f_{BE}(x,\mathbf{k})\delta_{sr},\label{eq:0th_MVSD}
\end{eqnarray}
which automatically yields the Bose-Einstein distribution $f_{BE}(x,\mathbf{k})$
when $s=r$. The first and second order results are,
\begin{eqnarray}
f_{sr}^{(1)}(x,\mathbf{k}) & = & \int d^{3}\mathbf{q}e^{-iq\cdot x}\left\{ -f_{BE}^{\prime}(x,\mathbf{p})\Xi_{++}+\frac{1}{2}f_{BE}^{\prime\prime}(x,\mathbf{p})\left[\beta\cdot p,\Xi_{++}\right]\right\} _{sr}\delta(\mathbf{q}),\label{eq:1st_MVSD_integral}
\end{eqnarray}
\begin{eqnarray}
f_{sr}^{(2)}(x,\mathbf{k}) & = & \int d^{3}\mathbf{q}e^{-iq\cdot x}\left\{ \frac{1}{2}f_{BE}^{\prime\prime}(x,\mathbf{p})\Xi_{++}^{2}-\frac{1}{6}f_{BE}^{\prime\prime\prime}(x,\mathbf{p})\Xi_{++}\left[\beta\cdot p,\Xi_{++}\right]\right.\nonumber \\
 &  & \left.-\frac{1}{3}f_{BE}^{\prime\prime\prime}(x,\mathbf{p})\left[\beta\cdot p,\Xi_{++}\right]\Xi_{++}+\frac{1}{8}f_{BE}^{\prime\prime\prime\prime}(x,\mathbf{p})[\beta\cdot p,\Xi_{++}]^{2}\right\} _{sr}\delta(\mathbf{q})\nonumber \\
 &  & +\widetilde{f}(x,\mathbf{k})\left(\Xi_{+-}\Xi_{-+}\right)_{sr},\label{eq:2rd_MVSD_integral}
\end{eqnarray}
where $f_{BE}^{\prime\cdots\prime}$ denotes the $n$-th derivative
$\partial^{n}f_{BE}/\partial^{n}(\beta\cdot k)$ with $n$ being the
number of primes, and $\widetilde{f}(x,\mathbf{k})$ is defined as
\begin{eqnarray}
\widetilde{f}(x,\mathbf{k}) & = & -\frac{1}{2\beta_{0}E_{\mathbf{k}}}\frac{e^{\beta\cdot k}}{(e^{\beta\cdot k}-1)^{2}}-\frac{1}{4\beta_{0}^{2}E_{\mathbf{k}}^{2}}\frac{e^{-\beta\cdot\tilde{k}}-e^{\beta\cdot k}}{(e^{\beta\cdot k}-1)(e^{-\beta\cdot\tilde{k}}-1)}.
\end{eqnarray}
In general, as mentioned, there will be additional second order terms
in the density operators $\hat{\varrho}_{B,C}$. The results presented
above do not account for these contributions.

After integrating over the $d^{3}\mathbf{q}$, Eqs.~(\ref{eq:1st_MVSD_integral},\ref{eq:2rd_MVSD_integral})
we obtain 
\begin{eqnarray}
f_{sr}^{(1)}(x,\mathbf{k}) & = & f_{BE}^{\prime}(x,\mathbf{k})\mathcal{F}_{sr},\label{eq:1st_MVSD}\\
f_{sr}^{(2)}(x,\mathbf{k}) & = & f_{2}(x,\mathbf{k})\delta_{sr}+\frac{1}{2}f_{BE}^{\prime\prime}(x,\mathbf{k})\sum_{s^{\prime}}\mathcal{F}_{ss^{\prime}}\mathcal{F}_{s^{\prime}r}+\widetilde{f}(x,\mathbf{k})\sum_{s^{\prime}}\Xi_{+-}^{ss^{\prime}}\Xi_{-+}^{s^{\prime}r},\label{eq:2rd_MVSD}
\end{eqnarray}
where $f_{2}(x,\mathbf{k})$ is defined as 
\begin{eqnarray}
f_{2}(x,\mathbf{k}) & = & (\xi_{\rho\sigma}-\omega_{\rho\sigma})(\xi_{\mu\nu}-\omega_{\mu\nu})\nonumber \\
 &  & \times\left[\frac{1}{12}f_{BE}^{\prime\prime\prime}(x,\mathbf{k})k^{\sigma}\beta_{\lambda}\left(g^{\rho\nu}-\frac{k^{\rho}}{E_{\mathbf{k}}}n^{\nu}\right)\left(g^{\mu\lambda}-\frac{k^{\mu}}{E_{\mathbf{k}}}n^{\lambda}\right)\right.\nonumber \\
 &  & +\frac{1}{8}f_{BE}^{\prime\prime}(x,\mathbf{k})\left(g^{\rho\nu}-\frac{k^{\rho}}{E_{\mathbf{k}}}n^{\nu}\right)\left(g^{\mu\sigma}-\frac{k^{\mu}}{E_{\mathbf{k}}}n^{\sigma}\right)\nonumber \\
 &  & \left.+\frac{1}{24}f_{BE}^{\prime\prime\prime}(x,\mathbf{k})k^{\sigma}k^{\nu}\beta_{0}\left(\frac{\Delta^{\rho\mu}}{E_{\mathbf{k}}}+\frac{\bar{k}^{\mu}\bar{k}^{\rho}}{E_{\mathbf{k}}^{3}}\right)\right],
\end{eqnarray}
with $\bar{k}^{\mu}=(0,\mathbf{k})$ and $\Delta^{\mu\nu}=g^{\mu\nu}-g^{\mu0}g^{\nu0}$,
and the functions $\mathcal{F}_{sr}$ corresponding to Belinfante
and canonical cases read 
\begin{eqnarray}
\mathcal{F}_{sr}^{(B)} & = & -i\omega_{\lambda\nu}\left(\frac{m}{E_{\mathbf{k}}+m}n^{\nu}\Psi_{sr}^{\lambda}+\frac{1}{2}\Psi_{sr}^{\nu\lambda}\right)+i\xi_{\lambda\nu}\frac{1}{E_{\mathbf{k}}}\frac{m}{E_{\mathbf{k}}+m}k^{\nu}\Psi_{sr}^{\lambda},\label{eq:1st_PDF_of_rhoB}\\
\mathcal{F}_{sr}^{(C)} & = & \frac{i}{2}\omega_{\lambda\nu}\frac{1}{E_{\mathbf{k}}}\frac{E_{\mathbf{k}}-m}{E_{\mathbf{k}}+m}k^{\nu}\Psi_{sr}^{\lambda}-\frac{i}{2}\xi_{\lambda\nu}\frac{1}{E_{\mathbf{k}}}\frac{E_{\mathbf{k}}-m}{E_{\mathbf{k}}+m}k^{\nu}\Psi_{sr}^{\lambda}\nonumber \\
 &  & -\frac{i}{2}\Omega_{\lambda\nu}\left(\Psi_{sr}^{\nu\lambda}+\frac{2m}{E_{\mathbf{k}}+m}n^{\nu}\Psi_{sr}^{\lambda}+\frac{1}{E_{\mathbf{k}}}\frac{E_{\mathbf{k}}-m}{E_{\mathbf{k}}+m}k^{\nu}\Psi_{sr}^{\lambda}\right),\label{eq:1st_PDF_of_rhoC}
\end{eqnarray}
where
\begin{eqnarray}
\Psi_{sr}^{\mu} & = & \frac{k\cdot\epsilon_{s}}{m}\epsilon_{r}^{\mu*}-\frac{k\cdot\epsilon_{r}^{*}}{m}\epsilon_{s}^{\mu}=\frac{1}{m}(0,-\mathbf{k}\times(\mathbf{e}_{s}\times\mathbf{e}_{r}^{*})),\nonumber \\
\Psi_{sr}^{\nu\lambda} & = & \epsilon_{r}^{\nu*}\epsilon_{s}^{\lambda}-\epsilon_{r}^{\lambda*}\epsilon_{s}^{\nu},
\end{eqnarray}
and $\epsilon_{s}^{\mu}=\epsilon_{s}^{\mu}(0)=(0,\mathbf{e}_{s})$
is the polarization vector in the particle's rest frame. After a lengthy
but straightforward calculation, we obtain expressions of $\sum_{s^{\prime}}\Xi_{+-}^{ss^{\prime}}\Xi_{-+}^{s^{\prime}s}$
for Belinfante and canonical cases, 
\begin{eqnarray}
\sum_{s^{\prime}}\Xi_{(B)+-}^{ss^{\prime}}\Xi_{(B)-+}^{s^{\prime}s} & = & \frac{1}{4}\xi^{\lambda\nu}\xi^{\rho\sigma}(\Xi_{(B)\beta\beta}^{ss})_{\lambda\nu,\rho\sigma},\label{eq:Sigma_B_01}\\
\sum_{s^{\prime}}\Xi_{(C)+-}^{ss^{\prime}}\Xi_{(C)-+}^{s^{\prime}s} & = & \frac{1}{4}(\xi^{\lambda\nu}-\omega^{\lambda\nu})(\xi^{\rho\sigma}-\omega^{\rho\sigma})(\Xi_{(C)\beta\beta}^{ss})_{\lambda\nu,\rho\sigma}+\frac{1}{4}\Omega^{\lambda\nu}\Omega^{\rho\sigma}(\Xi_{(C)\Omega\Omega}^{ss})_{\lambda\nu,\rho\sigma}\nonumber \\
 &  & +\frac{1}{4}(\xi^{\lambda\nu}-\omega^{\lambda\nu})\Omega^{\rho\sigma}\left[(\Xi_{(C)\beta\Omega}^{ss})_{\lambda\nu,\rho\sigma}+(\Xi_{(C)\beta\Omega}^{ss})_{\rho\sigma,\lambda\nu}^{*}\right],\label{eq:Sigma_C_01}
\end{eqnarray}
where the coefficients $\Xi^{ss}$ are given in Appendix~\ref{sec:Coefficients}. 


\section{Gradient expansion for $\rho^{00}$ }

\label{sec:rho00}

\subsection{First order results }

\label{subsec:1st_order}To the first order in the gradient expansion,
the spin density matrix $\rho_{sr}$ in Eq.~(\ref{eq:spin_density_matrix})
is given by 
\begin{eqnarray}
\rho_{sr}(x,\mathbf{k}) & = & \frac{f_{sr}^{(0)}+f_{sr}^{(1)}+\mathcal{O}(\partial^{2})}{\mathrm{Tr}\left(f^{(0)}+f^{(1)}\right)+\mathcal{O}(\partial^{2})}=\frac{1}{3}\delta_{sr}+\frac{1}{3f_{BE}}f_{sr}^{(1)}+\mathcal{O}(\partial^{2}).\label{eq:rho_sr_1st}
\end{eqnarray}
We can explicitly express the components of $\rho_{sr}$ as 
\begin{eqnarray}
\rho_{00}^{(B)}(x,\mathbf{k}) & = & \rho_{00}^{(C)}(x,\mathbf{k})+\mathcal{O}(\partial^{2}),\nonumber \\
\rho_{10}^{(B)}(x,\mathbf{k}) & = & \rho_{0,-1}^{(B)}(x,\mathbf{k})\nonumber \\
 & = & \frac{f_{BE}^{\prime}}{3f_{BE}}\left[\frac{1}{E_{\mathbf{k}}+m}\epsilon_{\mu\lambda\rho\sigma}n^{\lambda}k^{\rho}\epsilon_{+1}^{\sigma}\left(\omega^{\mu\nu}n_{\nu}-\xi^{\mu\nu}\frac{k_{\nu}}{E_{\mathbf{k}}}\right)\right.\nonumber \\
 &  & \left.+\frac{1}{2}\omega^{\mu\nu}\epsilon_{\mu\nu\rho\sigma}n^{\rho}\epsilon_{+1}^{\sigma}\right],\nonumber \\
\rho_{10}^{(C)}(x,\mathbf{k}) & = & \rho_{0,-1}^{(C)}(x,\mathbf{k})\nonumber \\
 & = & -\frac{f_{BE}^{\prime}}{6f_{BE}}\left[\frac{1}{E_{\mathbf{k}}+m}\epsilon_{\mu\lambda\rho\sigma}n^{\lambda}k^{\rho}\epsilon_{+1}^{\sigma}\left(\omega^{\mu\nu}k_{\nu}\frac{E_{\mathbf{k}}-m}{mE_{\mathbf{k}}}-\xi^{\mu\nu}k_{\nu}\frac{E_{\mathbf{k}}-m}{mE_{\mathbf{k}}}\right.\right.\nonumber \\
 &  & \times\left.\left.-\Omega^{\mu\nu}k_{\nu}\frac{E_{\mathbf{k}}-m}{mE_{\mathbf{k}}}-2\Omega^{\mu\nu}n_{\nu}\right)-\Omega^{\mu\nu}\epsilon_{\mu\nu\rho\sigma}n^{\rho}\epsilon_{+1}^{\sigma}\right]+\mathcal{O}(\partial^{2}),\nonumber \\
\rho_{1,-1}^{(B,C)}(x,\mathbf{k}) & = & \mathcal{O}(\partial^{2}).\label{eq:cross_section_01}
\end{eqnarray}
In global equilibrium, the Killing condition \citep{Becattini:2018duy}
gives $\xi^{\mu\nu}=0$ and $\Omega^{\mu\nu}=\omega^{\mu\nu}$. The
result in Eq.~(\ref{eq:cross_section_01}) reproduces that in Ref.~\citep{Gao:2023wwo},
and we have $\rho_{sr}^{(B)}=\rho_{sr}^{(C)}$ to the first order
in the gradient expansion. In local equilibrium, however, the thermal
shear stress tensor $\xi^{\mu\nu}$ is non-vanishing, so we have $\rho_{sr}^{(B)}\neq\rho_{sr}^{(C)}$
except $s=r=0$ to the first order.

A discussion about the result Eq.~(\ref{eq:cross_section_01}) is
needed. We notice that the unit vector $\mathbf{e}_{s}$ satisfies
the identity $\mathbf{e}_{s}^{*}=(-1)^{s}\mathbf{e}_{-s}$ which leads
to 
\begin{eqnarray}
\Psi_{sr}^{\lambda} & = & (-1)^{r}\frac{1}{m}(0,-\mathbf{k}\times(\mathbf{e}_{s}\times\mathbf{e}_{-r}))=-(-1)^{r+s}\Psi_{-r,-s}^{\lambda},\\
\Psi_{sr}^{\nu\lambda} & = & (-1)^{r}(\epsilon_{-r}^{\nu}\epsilon_{s}^{\lambda}-\epsilon_{-r}^{\lambda}\epsilon_{s}^{\nu})=-(-1)^{r+s}\Psi_{-r,-s}^{\nu\lambda}.\label{eq:relation_01}
\end{eqnarray}
Using the above relationship, we obtain
\begin{eqnarray}
\mathcal{F}_{i}^{sr} & = & -(-1)^{r+s}\mathcal{F}_{i}^{-r-s},
\end{eqnarray}
which further leads to 
\begin{eqnarray}
f_{sr}^{(1)}(x,\mathbf{k}) & = & -(-1)^{r+s}f_{-r,-s}^{(1)}(x,\mathbf{k}),\label{eq:1st_PDF_identity}
\end{eqnarray}
and 
\begin{eqnarray*}
\mathrm{Tr}f^{(1)} & = & f_{00}^{(1)}=f_{1,-1}^{(1)}=f_{-1,1}^{(1)}=0,\\
f_{10}^{(1)} & = & f_{0,-1}^{(1)}.
\end{eqnarray*}
These properties of $f_{sr}^{(1)}$ follows Eq.~(\ref{eq:cross_section_01}).


\subsection{Second order results}

To the second order in the gradient expansion the spin density matrix
reads 
\begin{eqnarray}
\rho_{sr}(x,\mathbf{k}) & = & \frac{f_{sr}^{(0)}+f_{sr}^{(1)}+f_{sr}^{(2)}+\mathcal{O}(\partial^{3})}{\mathrm{Tr}\left(f^{(0)}+f^{(1)}+f^{(2)}\right)+\mathcal{O}(\partial^{3})}\nonumber \\
 & = & \frac{1}{3}\delta^{sr}+\frac{1}{3f_{BE}}(f_{sr}^{(1)}+f_{sr}^{(2)})-\frac{1}{9f_{BE}^{2}}f_{sr}^{(0)}\mathrm{Tr}f^{(2)}+\mathcal{O}(\partial^{3}).\label{eq:spin_density_matrix-1}
\end{eqnarray}
We focus on the $\rho_{00}$ component of the spin density matrix.
Inserting Eq.~(\ref{eq:2rd_MVSD}) into Eq.~(\ref{eq:spin_density_matrix-1}),
we obtain 
\begin{eqnarray}
\rho_{00}^{(B)}(x,\mathbf{k}) & = & \frac{1}{3}-\frac{1}{18f_{BE}}f_{BE}^{\prime\prime}\Omega_{(B)}^{2}-\frac{1}{6f_{BE}}f_{BE}^{\prime\prime}(\Omega_{(B)}\cdot\epsilon_{0})^{2}\nonumber \\
 &  & +\frac{1}{36f_{BE}}\widetilde{f}\xi^{\lambda\nu}\xi^{\rho\sigma}\left[3(\Xi_{(B)\beta\beta}^{00})_{\lambda\nu,\rho\sigma}-\sum_{s}(\Xi_{(B)\beta\beta}^{ss})_{\lambda\nu,\rho\sigma}\right]+\mathcal{O}(\partial^{3}),\label{eq:rho00_B_2nd}
\end{eqnarray}
for the Belinfante case, and 
\begin{eqnarray}
\rho_{00}^{(C)}(x,\mathbf{k}) & = & \frac{1}{3}-\frac{1}{18f_{BE}}f_{BE}^{\prime\prime}\Omega_{(C)}^{2}-\frac{1}{6f_{BE}}f_{BE}^{\prime\prime}(\Omega_{(C)}\cdot\epsilon_{0})^{2}+\frac{1}{36f_{BE}}\widetilde{f}\nonumber \\
 &  & \times\left\{ (\xi^{\lambda\nu}-\omega^{\lambda\nu})(\xi^{\rho\sigma}-\omega^{\rho\sigma})\left[3(\Xi_{(C)\beta\beta}^{00})_{\lambda\nu,\rho\sigma}-\sum_{s}(\Xi_{(C)\beta\beta}^{ss})_{\lambda\nu,\rho\sigma}\right]\right.\nonumber \\
 &  & +(\xi^{\lambda\nu}-\omega^{\lambda\nu})\Omega^{\rho\sigma}\left[3(\Xi_{(C)\beta\Omega}^{00})_{\lambda\nu,\rho\sigma}+3(\Xi_{(C)\beta\Omega}^{00})_{\rho\sigma,\lambda\nu}^{*}\right.\nonumber \\
 &  & \left.-\sum_{s}(\Xi_{(C)\beta\Omega}^{ss})_{\lambda\nu,\rho\sigma}-\sum_{s}(\Xi_{(C)\beta\Omega}^{ss})_{\rho\sigma,\lambda\nu}^{*}\right]\nonumber \\
 &  & \left.+\Omega^{\lambda\nu}\Omega^{\rho\sigma}\left[3(\Xi_{(C)\Omega\Omega}^{00})_{\lambda\nu,\rho\sigma}-\sum_{s}(\Xi_{(C)\Omega\Omega}^{ss})_{\lambda\nu,\rho\sigma}\right]\right\} +\mathcal{O}(\partial^{3}),\label{eq:rho00_C_2nd}
\end{eqnarray}
for the canonical case. Here, we introduced the $\Omega_{(B)}$ and
$\Omega_{(C)}$ vectors for both Belinfante and canonical cases
\begin{eqnarray}
\Omega_{(B)}^{\sigma} & = & \frac{1}{E_{\mathbf{k}}+m}\epsilon^{\mu\lambda\rho\sigma}n_{\lambda}k_{\rho}\left(\omega_{\mu\nu}n^{\nu}-\xi_{\mu\nu}\frac{k^{\nu}}{E_{\mathbf{k}}}\right)+\frac{1}{2}\omega_{\mu\nu}\epsilon^{\mu\nu\rho\sigma}n_{\rho},\nonumber \\
\Omega_{(C)}^{\sigma} & = & -\frac{1}{2}\frac{1}{E_{\mathbf{k}}+m}\epsilon^{\mu\lambda\rho\sigma}n_{\lambda}k_{\rho}\left(\omega_{\mu\nu}k^{\nu}\frac{E_{\mathbf{k}}-m}{mE_{\mathbf{k}}}-\xi_{\mu\nu}k^{\nu}\frac{E_{\mathbf{k}}-m}{mE_{\mathbf{k}}}\right.\nonumber \\
 &  & \left.-\Omega_{\mu\nu}k^{\nu}\frac{E_{\mathbf{k}}-m}{mE_{\mathbf{k}}}-2\Omega_{\mu\nu}n^{\nu}\right)+\frac{1}{2}\Omega_{\mu\nu}\epsilon^{\mu\nu\rho\sigma}n_{\rho}.
\end{eqnarray}
We find that non-vanishing contributions to $\rho_{00}-1/3$ appear
in the second order of gradients. In the Belinfante case, the contributions
come from quadratic forms of thermal shear and vorticity tensors,
while in the canonical case there are extra terms of $(\partial^{\lambda}\beta^{\nu})(\partial^{\rho}\beta^{\sigma})$,
$(\partial^{\lambda}\beta^{\nu})\Omega^{\rho\sigma}$ and $\Omega^{\lambda\nu}\Omega^{\rho\sigma}$.


In the global equilibrium characterized by $\xi^{\mu\nu}=0$ and $\omega^{\mu\nu}=\Omega^{\mu\nu}$,
$\rho^{00}$ converges to the same value for both Belinfante and canonical
cases 
\begin{align}
\rho_{00}(x,\mathbf{k})= & \frac{1}{3}-\frac{1}{18f_{BE}}f_{BE}^{\prime\prime}\Omega^{2}-\frac{1}{6f_{BE}}f_{BE}^{\prime\prime}(\Omega\cdot\epsilon_{0})^{2}\nonumber \\
 & +\mathcal{O}(\partial^{3}),\label{eq:rho00_2nd_global}
\end{align}
where $\Omega^{\sigma}$ is given as 
\begin{align}
\Omega^{\sigma}\equiv & \Omega_{(B)}^{\sigma}=\Omega_{(C)}^{\sigma}\nonumber \\
= & \frac{1}{E_{\mathbf{k}}+m}\epsilon^{\mu\lambda\rho\sigma}n_{\lambda}k_{\rho}\omega_{\mu\nu}n^{\nu}+\frac{1}{2}\omega_{\mu\nu}\epsilon^{\mu\nu\rho\sigma}n_{\rho}.
\end{align}
In local equilibrium, the expressions for $\rho_{00}$ are different
for Belinfante and canonical cases, even when integrated over space.
The difference stems from distinct definitions of the density operator
$\hat{\varrho}$ in Eqs.~(\ref{eq:density_operator_B}) and (\ref{eq:density_operator_C}).
Notably, in the context of the spin polarization for $\Lambda$ and
$\overline{\Lambda}$ hyperons, $\hat{\varrho}$ is typically chosen
to be in the Belinfante form, as utilized in the modified Cooper-Frye
formula \citep{Becattini:2013fla,Fang:2016vpj}. However, we note
that the results of the canonical and other forms of density operators
are very different from those of the Belinfante form \citep{Buzzegoli:2021wlg}.
This discrepancy has led to discussions regarding the physical appropriateness
of these forms, particularly in how they align with experimental data.
The choice of a specific form is often made a posteriori, emphasizing
the need for further empirical validation \citep{Buzzegoli:2021wlg}.
So which form of the density operator has more reflection on the physical
reality remains an open question. This underscores the importance
of continued experimental and theoretical investigation to resolve
this issue.


\subsection{Another choice for MVSD \protect\label{sec:another=000020distribution=000020function}}

There is an alternative definition for the MVSD $f_{sr}$. From the
definition of the Wigner function in Eq.~(\ref{eq:Wigner_function_01}),
we can decompose the Wigner function into time-like and space-like
parts in momentum \citep{Becattini:2020sww}, 
\begin{eqnarray}
W^{\mu\nu}(x,k) & = & W_{+}^{\mu\nu}(x,k)+W_{-}^{\mu\nu}(x,k)+W_{S}^{\mu\nu}(x,k),
\end{eqnarray}
where $W_{S}^{\mu\nu}(x,k)$ is the space-like part and $W_{\pm}^{\mu\nu}(x,k)$
are the time-like parts (upper/lower sign corresponds to the positive/negative
energy branch respectively), defined as 
\begin{eqnarray}
W_{+}^{\mu\nu}(x,k) & = & W^{\mu\nu}(x,k)\theta(k_{0})\theta(k^{2}),\label{eq:Wigner_plus}\\
W_{-}^{\mu\nu}(x,k) & = & W^{\mu\nu}(x,k)\theta(-k_{0})\theta(k^{2}),\label{eq:Wigner_minus}\\
W_{S}^{\mu\nu}(x,k) & = & W^{\mu\nu}(x,k)\theta(-k^{2}).\label{eq:Wigner_S}
\end{eqnarray}
Different from Eq.~(\ref{eq:distribution_sr_a}), one can also introduce
an alternative MVSD (also see Refs.~\citep{DeGroot:1980dk,Becattini:2021lfq,Becattini:2020sww,Sheng:2021kfc}
for fermion's MVSD), 
\begin{eqnarray}
f_{sr}(x,k) & = & \frac{1}{2\pi}\epsilon_{\mu}^{r*}(\mathbf{k})\epsilon_{\nu}^{s}(\mathbf{k})W_{+}^{\mu\nu}(x,k)\nonumber \\
 & = & \frac{1}{(2\pi)^{3}}\int\frac{d^{3}q}{\sqrt{4E_{\mathbf{k}+\mathbf{q}/2}E_{\mathbf{k}-\mathbf{q}/2}}}\nonumber \\
 &  & \times\sum_{s^{\prime},r^{\prime}}e^{-iq\cdot x}\delta\left(k_{0}-\frac{E_{\mathbf{k}-\mathbf{q}/2}+E_{\mathbf{k}+\mathbf{q}/2}}{2}\right)\left\langle a_{\mathbf{k}-\mathbf{q}/2}^{s^{\prime}\dagger}a_{\mathbf{k}+\mathbf{q}/2}^{r^{\prime}}\right\rangle \nonumber \\
 &  & \times\epsilon_{\mu}^{r*}(\mathbf{k})\epsilon_{\nu}^{s}(\mathbf{k})\epsilon_{s^{\prime}}^{\nu*}\left(\mathbf{k}-\frac{\mathbf{q}}{2}\right)\epsilon_{r^{\prime}}^{\mu}\left(\mathbf{k}+\frac{\mathbf{q}}{2}\right).\label{eq:MVSD_Wigner_function}
\end{eqnarray}

We can make a gradient expansion following the same procedure as in
Eq.~(\ref{eq:gradient_expand_01}), we obtain 
\begin{eqnarray}
f_{sr}^{(0)}(x,k) & = & \delta(k^{2}-m^{2})\theta(k_{0})f_{BE}(x,\mathbf{k})\delta_{sr},\\
f_{sr}^{(1)}(x,k) & = & \delta(k^{2}-m^{2})\theta(k_{0})f_{BE}^{\prime}(x,\mathbf{k})\mathcal{F}_{sr},\nonumber \\
f_{sr}^{(2)}(x,k) & = & \delta(k^{2}-m^{2})\theta(k_{0})\left[f_{2}^{(+)}(x,\mathbf{k})\delta_{sr}+\frac{1}{2}f_{BE}^{\prime\prime}(x,\mathbf{k})\sum_{s^{\prime}}\mathcal{F}_{ss^{\prime}}\mathcal{F}_{s^{\prime}r}\right.\nonumber \\
 &  & +\widetilde{f}(x,\mathbf{k})\sum_{s^{\prime}}\Xi_{+-}^{ss^{\prime}}\Xi_{-+}^{s^{\prime}r}\nonumber \\
 &  & \left.-\frac{1}{4}f_{BE}^{\prime\prime}(x,\mathbf{k})(\xi_{\lambda\nu}-\omega_{\lambda\nu})(\xi_{\rho\sigma}-\omega_{\rho\sigma})\frac{k^{\nu}k^{\sigma}}{m^{2}}(\bar{\partial}_{k}^{\lambda}k^{\alpha})(\bar{\partial}_{k}^{\rho}k^{\beta})\epsilon_{\alpha}^{r*}(\mathbf{k})\epsilon_{\beta}^{s}(\mathbf{k})\right]\nonumber \\
 &  & -\frac{1}{16}\delta^{\prime}(k_{0}-E_{\mathbf{k}})f_{BE}^{\prime\prime}(x,\mathbf{k})(\xi_{\lambda\nu}-\omega_{\lambda\nu})(\xi_{\rho\sigma}-\omega_{\rho\sigma})k^{\sigma}k^{\nu}\left(\frac{\Delta^{\rho\lambda}}{E_{\mathbf{k}}^{2}}+\frac{\bar{k}^{\rho}\bar{k}^{\lambda}}{E_{\mathbf{k}}^{4}}\right)\delta_{sr},\label{eq:2rd_MVSD_Wigner}
\end{eqnarray}
with $\delta^{\prime}(x)=\partial\delta(x)/\partial x$ and
\begin{eqnarray}
f_{2}^{(+)}(x,\mathbf{k}) & = & f_{2}(x,\mathbf{k})-\frac{1}{4}f_{BE}^{\prime\prime}(\xi_{\lambda\nu}-\omega_{\lambda\nu})(\xi_{\rho\sigma}-\omega_{\rho\sigma})k^{\sigma}k^{\nu}\left(\frac{1}{2}\frac{\Delta^{\lambda\rho}}{E_{\mathbf{k}}^{2}}+\frac{\bar{k}^{\lambda}\bar{k}^{\rho}}{E_{\mathbf{k}}^{4}}\right),\\
\mathcal{F}_{sr}^{(B)} & = & -i\omega_{\lambda\nu}\epsilon_{r}^{\nu*}({\bf k})\epsilon_{s}^{\lambda}(\mathbf{k})+i\xi_{\lambda\nu}k^{\nu}\frac{1}{E_{\mathbf{k}}}\left[\epsilon_{r}^{0*}(\mathbf{k})\epsilon_{s}^{\lambda}(\mathbf{k})-\epsilon_{s}^{0}(\mathbf{k})\epsilon_{r}^{\lambda*}(\mathbf{k})\right],\label{eq:eq:1st_PDF_of_rhoB_Wigner}\\
\mathcal{F}_{sr}^{(C)} & = & -i\Omega_{\lambda\nu}\epsilon_{r}^{\nu*}({\bf k})\epsilon_{s}^{\lambda}(\mathbf{k})\nonumber \\
 &  & -i(\omega_{\lambda\nu}-\Omega_{\lambda\nu}-\xi_{\lambda\nu})k^{\nu}\frac{1}{2E_{\mathbf{k}}}\left[\epsilon_{r}^{0*}(\mathbf{k})\epsilon_{s}^{\lambda}(\mathbf{k})-\epsilon_{s}^{0}(\mathbf{k})\epsilon_{r}^{\lambda*}(\mathbf{k})\right].\label{eq:1st_PDF_of_rhoC_Wigner}
\end{eqnarray}
In the comparison with Eqs.~(\ref{eq:1st_PDF_of_rhoB}) and (\ref{eq:1st_PDF_of_rhoC}),
$\mathcal{F}_{sr}^{(B)}$ and $\mathcal{F}_{sr}^{(C)}$ in Eqs.~(\ref{eq:eq:1st_PDF_of_rhoB_Wigner})
and (\ref{eq:1st_PDF_of_rhoC_Wigner}) have an additional term 
\begin{equation}
\Delta\mathcal{F}_{sr}^{(B)}=\Delta\mathcal{F}_{sr}^{(C)}=(\partial_{\mu}\beta_{\nu})k^{\nu}[i\bar{\partial}_{k}^{\mu}\epsilon_{\beta}^{r*}({\bf k})]\epsilon_{s}^{\beta}({\bf k}),\label{eq:Delta_F}
\end{equation}
where $\overline{\partial}_{k}^{\mu}\equiv(0,-\boldsymbol{\nabla}_{k})$.
In global equilibrium, we observe that $\mathcal{F}_{sr}^{(B,C)}$
in Eqs.~(\ref{eq:eq:1st_PDF_of_rhoB_Wigner}) and (\ref{eq:1st_PDF_of_rhoC_Wigner})
do not depend on $n^{\mu}$, while $f_{sr}^{(2)}$ in Eq. (\ref{eq:2rd_MVSD_Wigner}),
$\mathcal{F}_{sr}^{(B,C)}$ in Eqs.~(\ref{eq:1st_PDF_of_rhoB}, \ref{eq:1st_PDF_of_rhoC})
and $f_{sr}^{(1,2)}$ in Eqs. (\ref{eq:1st_MVSD}) and (\ref{eq:2rd_MVSD})
do depend on $n^{\mu}$. According to Eq.~(\ref{eq:Delta_F}), the
extra term proportional to $\bar{\partial}_{k}^{\mu}\epsilon_{\beta}^{r*}({\bf k})$
cancels the terms related to $n^{\mu}$ in $\mathcal{F}_{sr}^{(B,C)}$
in Eqs.~(\ref{eq:1st_PDF_of_rhoB}, \ref{eq:1st_PDF_of_rhoC}).


From Eqs. (\ref{eq:spin_density_matrix}) and (\ref{eq:spin_density_matrix-1})
we can obtain the spin density matrix as the normalized MVSD in Eqs.~(\ref{eq:2rd_MVSD_Wigner}),
\begin{eqnarray}
\rho_{00}^{(B)}(x,\mathbf{k}) & = & \frac{1}{3}-\frac{1}{18f_{BE}}f_{BE}^{\prime\prime}\Omega_{(B)}^{2}-\frac{1}{6f_{BE}}f_{BE}^{\prime\prime}(\Omega_{(B)}\cdot\epsilon_{0})^{2}\nonumber \\
 &  & -\frac{1}{36}f_{BE}^{\prime\prime}(\xi_{\lambda\nu}-\omega_{\lambda\nu})(\xi_{\rho\sigma}-\omega_{\rho\sigma})\Xi_{(+)}^{\lambda\nu,\rho\sigma}\nonumber \\
 &  & +\frac{1}{36f_{BE}}\widetilde{f}\xi^{\lambda\nu}\xi^{\rho\sigma}\left[3(\Xi_{(B)\beta\beta}^{ss})_{\lambda\nu,\rho\sigma}-\sum_{s}(\Xi_{(B)\beta\beta}^{ss})_{\lambda\nu,\rho\sigma}\right]+\mathcal{O}(\partial^{3}),\label{eq:another-rho-b}
\end{eqnarray}
for the Belinfante case, and 
\begin{eqnarray}
\rho_{00}^{(C)}(x,\mathbf{k}) & = & \frac{1}{3}-\frac{1}{18f_{BE}}f_{BE}^{\prime\prime}\Omega_{(C)}^{2}-\frac{1}{6f_{BE}}f_{BE}^{\prime\prime}(\Omega_{(C)}\cdot\epsilon_{0})^{2}\nonumber \\
 &  & -\frac{1}{36}f_{BE}^{\prime\prime}(\xi_{\lambda\nu}-\omega_{\lambda\nu})(\xi_{\rho\sigma}-\omega_{\rho\sigma})\Xi_{(+)}^{\lambda\nu,\rho\sigma}\nonumber \\
 &  & +\frac{1}{36f_{BE}}\widetilde{f}\left\{ (\xi^{\lambda\nu}-\omega^{\lambda\nu})(\xi^{\rho\sigma}-\omega^{\rho\sigma})\left[3(\Xi_{(C)\beta\beta}^{00})_{\lambda\nu,\rho\sigma}-\sum_{s}(\Xi_{(C)\beta\beta}^{ss})_{\lambda\nu,\rho\sigma}\right]\right.\nonumber \\
 &  & +(\xi^{\lambda\nu}-\omega^{\lambda\nu})\Omega^{\rho\sigma}\left[3(\Xi_{(C)\beta\Omega}^{00})_{\lambda\nu,\rho\sigma}+3(\Xi_{(C)\beta\Omega}^{00})_{\rho\sigma,\lambda\nu}^{*}\right.\nonumber \\
 &  & \left.-\sum_{s}(\Xi_{(C)\beta\Omega}^{ss})_{\lambda\nu,\rho\sigma}-\sum_{s}(\Xi_{(C)\beta\Omega}^{ss})_{\rho\sigma,\lambda\nu}^{*}\right]\nonumber \\
 &  & \left.+\Omega^{\lambda\nu}\Omega^{\rho\sigma}\left[3(\Xi_{(C)\Omega\Omega}^{00})_{\lambda\nu,\rho\sigma}-\sum_{s}(\Xi_{(C)\Omega\Omega}^{ss})_{\lambda\nu,\rho\sigma}\right]\right\} +\mathcal{O}(\partial^{3}),\label{eq:another-rho-c}
\end{eqnarray}
for the canonical case, where we defined 
\begin{eqnarray}
\Omega_{(B)}^{\sigma} & = & \frac{1}{m}\epsilon^{\mu\lambda\rho\sigma}n_{\lambda}k_{\rho}\left(\omega_{\mu\nu}n^{\nu}\frac{m}{E_{\mathbf{k}}+m}+\omega_{\mu\nu}\frac{k^{\nu}}{E_{\mathbf{k}}+m}-\xi_{\mu\nu}\frac{k^{\nu}}{E_{\mathbf{k}}}\right)\nonumber \\
 &  & +\frac{1}{2}\omega_{\mu\nu}\epsilon^{\mu\nu\rho\sigma}n_{\rho},\\
\Omega_{(C)}^{\sigma} & = & -\frac{1}{2m}\epsilon^{\mu\lambda\rho\sigma}n_{\lambda}k_{\rho}\left(\xi_{\mu\nu}\frac{k^{\nu}}{E_{\mathbf{k}}}-\omega_{\mu\nu}\frac{k^{\nu}}{E_{\mathbf{k}}}-\Omega_{\mu\nu}\frac{k^{\nu}}{E_{\mathbf{k}}}\frac{E_{\mathbf{k}}-m}{E_{\mathbf{k}}+m}\right.\nonumber \\
 &  & \left.-\frac{2m}{E_{\mathbf{k}}+m}\Omega^{\mu\nu}n_{\nu}\right)+\frac{1}{2}\Omega_{\mu\nu}\epsilon^{\mu\nu\rho\sigma}n_{\rho},\\
\Xi_{(+)}^{\lambda\nu,\rho\sigma} & = & 3\frac{k^{\nu}k^{\sigma}}{m^{2}}\left[\epsilon_{0}^{\lambda*}\epsilon_{0}^{\rho}+\frac{m}{E_{\mathbf{k}}(E_{\mathbf{k}}+m)}\left(\frac{k\cdot\epsilon_{0}^{*}}{m}\epsilon_{0}^{\rho}\bar{k}^{\lambda}+\frac{k\cdot\epsilon_{0}}{m}\epsilon_{0}^{\lambda*}\bar{k}^{\rho}\right)\right.\nonumber \\
 &  & \left.+\frac{m^{2}}{E_{\mathbf{k}}^{2}(E_{\mathbf{k}}+m)^{2}}\frac{k\cdot\epsilon_{0}^{*}}{m}\frac{k\cdot\epsilon_{0}}{m}\bar{k}^{\lambda}\bar{k}^{\rho}\right]+\frac{k^{\nu}k^{\sigma}}{m^{2}}\left(\Delta^{\lambda\rho}+\frac{\bar{k}^{\lambda}\bar{k}^{\rho}}{E_{\mathbf{k}}^{2}}\right).
\end{eqnarray}
Here, we simply drop $\delta(k^{2}-m^{2})$ in the numerator and denominator
in the definition of the spin density matrix (\ref{eq:spin_density_matrix-1})
and take $k_{0}=E_{\mathbf{k}}$.


We can define the momentum dependent spin density matrix as, 
\begin{eqnarray}
\rho_{sr}(\mathbf{k}) & = & \frac{\int d^{3}xf_{sr}(x,k)}{\int d^{3}x\mathrm{Tr}f(x,k)}=\frac{\delta(k^{2}-m^{2})\textrm{Tr}(\hat{\varrho}a_{\mathbf{k}}^{s\dagger}a_{\mathbf{k}}^{r})}{\sum_{s^{\prime}}\delta(k^{2}-m^{2})\textrm{Tr}(\hat{\varrho}a_{\mathbf{k}}^{s^{\prime}\dagger}a_{\mathbf{k}}^{s^{\prime}})}\nonumber \\
 & = & \frac{\textrm{Tr}(\hat{\varrho}a_{\mathbf{k}}^{s\dagger}a_{\mathbf{k}}^{r})}{\sum_{s^{\prime}}\textrm{Tr}(\hat{\varrho}a_{\mathbf{k}}^{s^{\prime}\dagger}a_{\mathbf{k}}^{s^{\prime}})},\label{eq:density_matirx_k_Wigner}
\end{eqnarray}
where in the third equality, $\delta(k^{2}-m^{2})$ functions cancel
in the numerator and denominator and $k_{0}=E_{\mathbf{k}}$ is taken.
Interestingly, although $\rho_{00}^{(B,C)}(x,\mathbf{k})$ in Eqs.
(\ref{eq:another-rho-b}) and (\ref{eq:another-rho-c}) differ from
$\rho_{00}^{(B,C)}(x,\mathbf{k})$ in Eqs.~(\ref{eq:rho00_B_2nd},
\ref{eq:rho00_C_2nd}), the momentum dependent spin density matrix
is the same for both choices of MVSDs.



\section{Time reversal property of spin alignment induced by shear stress
tensor }

\label{sec:dissipative}In this section, we discuss the time reversal
property of the spin alignment induced by the shear stress tensor,
which is related to the question if the spin alignment induced by
the shear stress tensor dissipative.

If a transport coefficient in hydrodynamics is even under the time
reversal transformation (T-even), the transport effect is called non-dissipative,
while it is called dissipative if the corresponding transport coefficient
is T-odd \citep{Kharzeev:2011ds,Chen:2016xtg}. For example, the chiral
magnetic effect is described by $\mathbf{j}=\xi_{B}\mathbf{B}$ with
$\xi_{B}$ being the transport coefficient and is non-dissipative
as it does not lead to entropy production \citep{Son:2009tf,Pu:2010as}.
This can be seen by the time reversal transformation of the current
and magnetic field, $\mathbf{j}\rightarrow-\mathbf{j}$ and $\mathbf{B}\rightarrow-\mathbf{B}$,
so the coefficient $\xi_{B}$ is T-even.


We consider the spin alignment induced by the shear stress tensor
in this work. From Eq.~(\ref{eq:MVSD_Wigner_function}), we have
\begin{equation}
\rho_{sr}(x,\mathbf{k})\sim\epsilon_{\mu}^{r*}(\mathbf{k})\epsilon_{\nu}^{s}(\mathbf{k})W^{\mu\nu}.
\end{equation}
Given $W^{\mu\nu}$ being a symmetric rank-2 tensor, it can be associated
with other symmetric tensors. In this context, we consider following
terms 
\begin{eqnarray}
\epsilon_{\mu}^{r*}(\mathbf{k})\epsilon_{\nu}^{s}(\mathbf{k})W^{\mu\nu} & \sim & a\xi^{\mu\nu}\epsilon_{\mu}^{r*}(\mathbf{k})\epsilon_{\nu}^{s}(\mathbf{k})\nonumber \\
 &  & +b\left(\xi^{\mu\lambda}\xi_{\lambda}^{\ \nu}+\xi^{\mu\lambda}\xi_{\lambda}^{\ \nu}\right)\epsilon_{\mu}^{r*}(\mathbf{k})\epsilon_{\nu}^{s}(\mathbf{k})+\mathcal{O}(\xi_{\mu\nu}^{3}),
\end{eqnarray}
where $a$ and $b$ are interpreted as transport coefficients. Under
the time reversal transformation, the polarization vector transforms
as 
\begin{eqnarray}
\epsilon_{0}^{s}(\mathbf{k}) & \rightarrow & (-1)^{s+1}\epsilon_{0}^{-s}(-\mathbf{k}),\quad\epsilon_{i}^{s}(\mathbf{k})\rightarrow(-1)^{s}\epsilon_{i}^{-s}(-\mathbf{k}),
\end{eqnarray}
which leads to the time reversal transformation of the spin density
matrix $\rho_{sr}$ as 
\begin{eqnarray}
\rho_{sr}(t,\mathbf{x},\mathbf{k}) & \rightarrow & (-1)^{s+r}\rho_{-r,-s}(-t,\mathbf{x},-\mathbf{k}).
\end{eqnarray}
Additional related quantities under the time reversal transformation
are detailed in Table~\ref{tab:Quantities}. From this analysis,
we find that the transport coefficient $a$ is T-odd while $b$ is
T-even for $\rho_{00}$. Therefore, following this analysis, it implies
that the spin alignment induced by the shear stress tensor in the
first order of the gradient expansion is dissipative.

\begin{center}
\begin{table}
\caption{The time reversal transformation (TRT) for components of the shear
stress tensor and spin density matrix. \protect\label{tab:Quantities}}

\centering{}%
\begin{tabular}{|c|c|c|c|c|c|c|}
\hline 
terms & $\partial_{0}$ & $\beta_{0}$ & $\xi_{00}$ & $\xi_{0i}$ & $\xi_{ij}$ & $(-1)^{s+r}\rho_{sr}(t,\mathbf{x},\mathbf{k})$\tabularnewline
\hline 
TRT & $-\partial_{0}$ & $+\beta_{0}$ & $-\xi_{00}$ & $+\xi_{0i}$ & $-\xi_{ij}$ & $\rho_{-r,-s}(-t,\mathbf{x},-\mathbf{k})$\tabularnewline
\hline 
terms & $\partial_{i}$ & $\beta_{i}$ & $(-1)^{s+r}\epsilon_{0}^{r*}(\mathbf{k})\epsilon_{0}^{s}(\mathbf{k})$ & $(-1)^{s+r}\epsilon_{0}^{r*}(\mathbf{k})\epsilon_{i}^{s}(\mathbf{k})$ & $(-1)^{s+r}\epsilon_{i}^{r*}(\mathbf{k})\epsilon_{j}^{s}(\mathbf{k})$ & \tabularnewline
\hline 
TRT & $+\partial_{i}$ & $-\beta_{i}$ & $+\epsilon_{0}^{-r*}(-\mathbf{k})\epsilon_{0}^{-s}(-\mathbf{k})$ & $-\epsilon_{0}^{-r*}(-\mathbf{k})\epsilon_{i}^{-s}(-\mathbf{k})$ & $+\epsilon_{i}^{-r*}(-\mathbf{k})\epsilon_{j}^{-s}(-\mathbf{k})$ & \tabularnewline
\hline 
\end{tabular}
\end{table}
\par\end{center}


The density operator $\hat{\varrho}_{\textrm{R}}$ for a real system
in local equilibrium differs from $\hat{\varrho}$ introduced in Eqs.~(\ref{eq:density_operator_B})
and (\ref{eq:density_operator_C}). From $\hat{\varrho}_{\textrm{R}}$
and $\hat{\varrho}$ we can obtain statistical average values of the
energy momentum tensor $T_{\textrm{R}}^{\mu\nu}\equiv\text{Tr}(\hat{\varrho}_{\textrm{R}}\hat{T}^{\mu\nu})$
and $T^{\mu\nu}\equiv\text{Tr}(\hat{\varrho}\hat{T}^{\mu\nu})$ whcih
are related by \citep{van1982maximum,Becattini:2019dxo}, 
\begin{equation}
n_{\mu}T_{\textrm{R}}^{\mu\nu}=n_{\mu}T^{\mu\nu},
\end{equation}
where $n_{\mu}$ is the unit vector perpendicular to the hypersurface
$\Sigma$ chosen in the density operator $\hat{\varrho}$. It is important
to note that $T_{\textrm{R}}^{\mu\nu}$ is not necessarily equal to
$T^{\mu\nu}$, which is computed from the density operators in Eqs.~(\ref{eq:density_operator_B})
and (\ref{eq:density_operator_C}). Following the Refs.~\citep{van1982maximum,Becattini:2019dxo},
one can demonstrate that the entropy production rate is given by 
\begin{eqnarray}
\partial_{\mu}s^{\mu} & = & (T_{\textrm{R}}^{\mu\nu}-T^{\mu\nu})\partial_{\mu}\beta_{\nu}.
\end{eqnarray}
When $\hat{\varrho}=\hat{\varrho}_{\textrm{R}}$, we have $T_{\textrm{R}}^{\mu\nu}=T^{\mu\nu}$,
resulting in conserved entropy. Consequently, any thermal quantity
$\textrm{Tr }(\hat{\varrho}\hat{O})$ computed using the density operator
$\hat{\varrho}$ in Eqs.~(\ref{eq:density_operator_B}) and (\ref{eq:density_operator_C})
are non-dissipative. While the real ensemble averaged quantity $\textrm{Tr }(\hat{\varrho}_{\textrm{R}}\hat{O})$
can be dissipative which can be computed using the Kubo formula \citep{zubarev1981modern,Becattini:2019dxo},
\begin{eqnarray}
 &  & \text{Tr}[\hat{\varrho}_{\textrm{R}}\hat{O}(x)]-\text{Tr}[\hat{\varrho}\hat{O}(x)]\nonumber \\
 & = & \partial_{\mu}\beta_{\nu}(x)\lim_{k\to0}\textrm{Im}T\int_{t_{0}}^{t}d^{4}x^{\prime}\int_{t_{0}}^{t^{\prime}}d\theta\left\langle [\hat{T}^{\mu\nu}(\theta,\mathbf{x}^{\prime}),\hat{O}(x)]\right\rangle _{\beta(x)}e^{-ik\cdot(x^{\prime}-x)}.
\end{eqnarray}
For the spin alignment induced by the shear stress tensor computed
using the above formula, we refer the readers to Ref.~\citep{Dong:2023cng}.



\section{Summary }

\label{sec:summary}We have computed the spin density matrix element
$\rho_{00}$ up to the second order of gradient expansion at local
equilibrium using Zubarev's approach. The spin density matrix is obtained
through the matrix valued spin dependent distribution function $f_{sr}$.
It is defined as the ensemble average on the density operators $\hat{\varrho}$
which depend on the choice of the energy-momentum and spin tensors
or pseudo-gauge. We choose two forms of the energy-momentum and spin
tensors: Belinfante and canonical ones. Then we obtain $f_{sr}$ up
to the second order by making the gradient expansion for $\hat{\varrho}$.


The results for $\rho_{00}$ to the first order are given in Eqs.~(\ref{eq:cross_section_01}).
We find $\rho_{00}=1/3$, meaning that there is no contribution from
the thermal vorticity and shear stress tensor to the first order.
The off-diagonal elements of the spin density matrix are nonzero.


The results for $\rho_{00}$ in the second order are presented in
Eqs.~(\ref{eq:rho00_B_2nd}) and (\ref{eq:rho00_C_2nd}) with non-vanishing
contributions from both the thermal vorticity and shear stress tensors.
The exact expression for $\rho_{00}$ depends on the pseudo-gauge,
so $\rho_{00}$ are different in the Belinfante and canonical cases.
In global equilibrium with Killing conditions, $\rho_{00}$ in both
cases converge to a unique expression as shown in Eq.~(\ref{eq:rho00_2nd_global}).
The choice of the pseudo-gauge is a posteriori and should be validated
by experimental data.


The definition of $f_{sr}$ is also not unique. We discussed an alternative
choice of $f_{sr}$ in Eq.~(\ref{eq:MVSD_Wigner_function}). The
expressions for $f_{sr}$ up to the second order of gradient expansion
are given in Eq.~(\ref{eq:2rd_MVSD_Wigner}). Although the spin density
matrix are different from the previous one, they give the same integrated
result over the spatial coordinate, as shown in Eq.~(\ref{eq:density_matirx_k_Wigner}).


We finally discussed the time reversal property of the spin alignment
induced by the shear stress tensor, which is related to the question
if the spin alignment induced by the shear stress tensor dissipative.
The effective transport coefficient for the spin alignment induced
by the thermal shear stress tensor is T-odd in the first order, impling
that the first order effect is dissipative. For the ensemble defined
by the density operators in Eq. (\ref{eq:rho_B_gradient_expation})
and (\ref{eq:rho_c_gradient_expation-1}), the contribution to the
spin alignment from the thermal shear stress tensor only appears in
the second order, since such density operators can only give non-dissipative
effects.


\begin{acknowledgments}
We would like to thank Francesco Becattini, Xu-Guang Huang, Xin-Li
Sheng and Zong-Hua Zhang for helpful discussion. This work is supported
in part by the National Key Research and Development Program of China
under Contract No. 2022YFA1605500, by the Chinese Academy of Sciences
(CAS) under Grants No. YSBR-088 and by National Nature Science Foundation
of China (NSFC) under Grants No. 12075235, 12135011, 12175123 and
12147101.
\end{acknowledgments}

\paragraph{Note added. }

Recently, we were informed of Ref. \citep{Zhang:2024} which works
on a similar topic and appeared on arXiv on the same day.

\appendix

\section{Derivations of Eqs.~(\ref{eq:commutation_A+})-(\ref{eq:rho_a_rho})}

\label{sec:Detail-of-commutation}From Eqs.~(\ref{eq:vector_field}),
(\ref{eq:R3_spin_tensor}), (\ref{eq:relation_TB_TC}) and (\ref{eq:hat_P}),
we obtain 

\begin{eqnarray}
\hat{P}^{\nu} & = & \frac{1}{2}\int d^{3}yn_{\mu}\int\frac{d^{3}p}{(2\pi)^{3}}\int\frac{d^{3}q}{(2\pi)^{3}}\sum_{s,r}\nonumber \\
 &  & \times\left[\Phi_{(1)+-}^{sr;\mu\nu}(p,q)a_{{\bf p}}^{s\dagger}a_{{\bf q}}^{r}e^{i(p-q)\cdot y}+\Phi_{(1)-+}^{sr;\mu\nu}(p,q)a_{{\bf p}}^{s}a_{{\bf q}}^{r\dagger}e^{-i(p-q)\cdot y}\right.\nonumber \\
 &  & \left.-\Phi_{(1)--}^{sr;\mu\nu}(p,q)a_{{\bf p}}^{s}a_{{\bf q}}^{r}e^{-i(p+q)\cdot y}-\Phi_{(1)++}^{sr;\mu\nu}(p,q)a_{{\bf p}}^{s\dagger}a_{{\bf q}}^{r\dagger}e^{i(p+q)\cdot y}\right],\label{eq:P_before}
\end{eqnarray}
where the auxiliary functions $\Phi_{(1)}$ are 

\begin{eqnarray}
\Phi_{(1)+-}^{sr;\mu\nu}(p,q) & = & \left[\Phi_{(1)-+}^{sr;\mu\nu}(p,q)\right]^{*}\nonumber \\
 & = & \frac{1}{\sqrt{2E_{\mathbf{p}}}}\frac{1}{\sqrt{2E_{\mathbf{q}}}}\left\{ \left[(p\cdot q-m^{2})g^{\mu\nu}-2p^{\mu}q^{\nu}\right]\left[\epsilon_{s}^{*}({\bf p})\cdot\epsilon_{r}({\bf q})\right]\right.\nonumber \\
 &  & \left.-g^{\mu\nu}\left[q\cdot\epsilon_{s}^{*}({\bf p})\right]\left[p\cdot\epsilon_{r}({\bf q})\right]+2q^{\nu}\epsilon_{s}^{\mu*}({\bf p})\left[p\cdot\epsilon_{r}({\bf q})\right]\right\} ,\label{eq:Phi_1pm}
\end{eqnarray}

\begin{eqnarray}
\Phi_{(1)--}^{sr;\mu\nu}(p,q) & = & \left[\Phi_{(1)++}^{sr;\mu\nu}(p,q)\right]^{*}\nonumber \\
 & = & \frac{1}{\sqrt{2E_{\mathbf{p}}}}\frac{1}{\sqrt{2E_{\mathbf{q}}}}\left\{ \left[(p\cdot q+m^{2})g^{\mu\nu}-2p^{\mu}q^{\nu}\right]\left[\epsilon_{s}({\bf p})\cdot\epsilon_{r}({\bf q})\right]\right.\nonumber \\
 &  & \left.-g^{\mu\nu}\left[q\cdot\epsilon_{s}({\bf p})\right]\left[p\cdot\epsilon_{r}({\bf q})\right]+2q^{\nu}\epsilon_{s}^{\mu}({\bf p})\left[p\cdot\epsilon_{r}({\bf q})\right]\right\} .\label{eq:Phi_1mm}
\end{eqnarray}
After integrating over $y$ and $q$ in Eq.~(\ref{eq:P_before})
and replacing $p$ with $q$ as the integral variable, we obtain 
\begin{eqnarray}
\hat{P}^{\nu} & = & \frac{1}{2}n_{\mu}\int\frac{d^{3}q}{(2\pi)^{3}}\sum_{s,r}[\Phi_{(1)+-}^{sr;\mu\nu}(q,q)a_{{\bf q}}^{s\dagger}a_{{\bf q}}^{r}+\Phi_{(1)-+}^{sr;\mu\nu}(q,q)a_{{\bf q}}^{s}a_{{\bf q}}^{r\dagger}\nonumber \\
 &  & -\Phi_{(1)--}^{sr;\mu\nu}(q,\tilde{q})a_{{\bf q}}^{s}a_{-{\bf q}}^{r}e^{-2iE_{\mathbf{q}}t_{y}}-\Phi_{(1)++}^{sr;\mu\nu}(q,\tilde{q})a_{{\bf q}}^{s\dagger}a_{-{\bf q}}^{r\dagger}e^{2iE_{\mathbf{q}}t_{y}}],
\end{eqnarray}
where $\tilde{q}=(E_{\mathbf{q}},-\mathbf{q})$. The commutators are 

\begin{eqnarray}
[\hat{P}^{\nu},a_{\mathbf{p}}^{s\dagger}] & = & \frac{1}{2}n_{\mu}\sum_{r}\left[\Phi_{(1)+-}^{rs;\mu\nu}(p,p)+\Phi_{(1)-+}^{sr;\mu\nu}(p,p)\right]a_{{\bf p}}^{r\dagger}\nonumber \\
 &  & -\frac{1}{2}n_{\mu}\sum_{r}\left[\Phi_{(1)--}^{rs;\mu\nu}(\tilde{p},p)+\Phi_{(1)--}^{sr;\mu\nu}(p,\tilde{p})\right]a_{-{\bf p}}^{r}e^{-2iE_{\mathbf{p}}t_{y}}\nonumber \\
 & = & p^{\nu}a_{{\bf p}}^{s\dagger},\nonumber \\{}
[\hat{P}^{\nu},a_{-\mathbf{p}}^{s}] & = & -\frac{1}{2}n_{\mu}\sum_{r}\left[\Phi_{(1)+-}^{rs;\mu\nu}(\tilde{p},\tilde{p})+\Phi_{(1)-+}^{sr;\mu\nu}(\tilde{p},\tilde{p})\right]a_{-{\bf p}}^{r}\nonumber \\
 &  & +\frac{1}{2}n_{\mu}\sum_{r}\left[\Phi_{(1)++}^{rs;\mu\nu}(p,\tilde{p})+\Phi_{(1)++}^{sr;\mu\nu}(\tilde{p},p)\right]a_{{\bf p}}^{r\dagger}e^{2iE_{\mathbf{p}}t_{y}}\nonumber \\
 & = & -\tilde{p}^{\nu}a_{-{\bf p}}^{s},
\end{eqnarray}
where we have used following identities 
\begin{eqnarray}
n_{\mu}\left[\Phi_{(1)+-}^{rs;\mu\nu}(p,p)+\Phi_{(1)-+}^{sr;\mu\nu}(p,p)\right] & = & 2p^{\nu}\delta_{sr},\nonumber \\
n_{\mu}\left[\Phi_{(1)--}^{rs;\mu\nu}(\tilde{p},p)+\Phi_{(1)--}^{sr;\mu\nu}(p,\tilde{p})\right] & = & n_{\mu}\left[\Phi_{(1)++}^{rs;\mu\nu}(p,\tilde{p})+\Phi_{(1)++}^{sr;\mu\nu}(\tilde{p},p)\right]=0.
\end{eqnarray}

Analogy to above formula, we obtain expressions for other operators.
From Eqs. (\ref{eq:hat_P}), (\ref{eq:hat_L}) and (\ref{eq:P_before})
we obtain
\begin{eqnarray}
\hat{L}^{\lambda\nu} & = & -x^{\lambda}\hat{P}^{\nu}+\frac{1}{2}\int d^{3}yy^{\lambda}n_{\mu}\int\frac{d^{3}p}{(2\pi)^{3}}\int\frac{d^{3}q}{(2\pi)^{3}}\sum_{s,r}\nonumber \\
 &  & \times\left[\Phi_{(1)+-}^{sr;\mu\nu}(p,q)a_{{\bf p}}^{s\dagger}a_{{\bf q}}^{r}e^{i(p-q)\cdot y}+\Phi_{(1)-+}^{sr;\mu\nu}(p,q)a_{{\bf p}}^{s}a_{{\bf q}}^{r\dagger}e^{-i(p-q)\cdot y}\right.\nonumber \\
 &  & \left.-\Phi_{(1)--}^{sr;\mu\nu}(p,q)a_{{\bf p}}^{s}a_{{\bf q}}^{r}e^{-i(p+q)\cdot y}-\Phi_{(1)++}^{sr;\mu\nu}(p,q)a_{{\bf p}}^{s\dagger}a_{{\bf q}}^{r\dagger}e^{i(p+q)\cdot y}\right].\label{eq:oam}
\end{eqnarray}
From Eqs. (\ref{eq:R3_spin_tensor}), (\ref{eq:hat_S_B}) and (\ref{eq:hat_S_C}),
we obtain
\begin{eqnarray}
\hat{S}_{B}^{\lambda\nu} & = & -2\int d^{3}yn_{\mu}\int\frac{d^{3}p}{(2\pi)^{3}}\int\frac{d^{3}q}{(2\pi)^{3}}\sum_{s,r}\nonumber \\
 &  & \times\left[\Phi_{(2)+-}^{sr;\mu\lambda\nu}(p,q)a_{\mathbf{p}}^{s\dagger}a_{{\bf q}}^{r}e^{i(p-q)\cdot y}+\Phi_{(2)-+}^{sr;\mu\lambda\nu}(p,q)a_{{\bf p}}^{s}a_{{\bf q}}^{r\dagger}e^{-i(p-q)\cdot y}\right.\nonumber \\
 &  & \left.-\Phi_{(2)--}^{sr;\mu\lambda\nu}(p,q)a_{\mathbf{p}}^{s}a_{{\bf q}}^{r}e^{-i(p+q)\cdot y}-\Phi_{(2)++}^{sr;\mu\lambda\nu}(p,q)a_{\mathbf{p}}^{s\dagger}a_{{\bf q}}^{r\dagger}e^{i(p+q)\cdot y}\right],\label{eq:SB}
\end{eqnarray}
\begin{eqnarray}
\hat{S}_{C}^{\lambda\nu} & = & -\int d^{3}yn_{\mu}\int\frac{d^{3}p}{(2\pi)^{3}}\int\frac{d^{3}q}{(2\pi)^{3}}\sum_{s,r}\nonumber \\
 &  & \times\left\{ \left[\Phi_{(2)+-}^{sr;\mu\lambda\nu}(p,q)-\Phi_{(2)+-}^{sr;\mu\nu\lambda}(p,q)\right]a_{\mathbf{p}}^{s\dagger}a_{{\bf q}}^{r}e^{i(p-q)\cdot y}\right.\nonumber \\
 &  & +\left[\Phi_{(2)-+}^{sr;\mu\lambda\nu}(p,q)-\Phi_{(2)-+}^{sr;\mu\nu\lambda}(p,q)\right]a_{\mathbf{p}}^{s}a_{{\bf q}}^{r\dagger}e^{-i(p-q)\cdot y}\nonumber \\
 &  & -\left[\Phi_{(2)--}^{sr;\mu\lambda\nu}(p,q)-\Phi_{(2)--}^{sr;\mu\nu\lambda}(p,q)\right]a_{\mathbf{p}}^{s}a_{{\bf q}}^{r}e^{-i(p+q)\cdot y}\nonumber \\
 &  & \left.-\left[\Phi_{(2)++}^{sr;\mu\lambda\nu}(p,q)-\Phi_{(2)++}^{sr;\mu\nu\lambda}(p,q)\right]a_{\mathbf{p}}^{s\dagger}a_{{\bf q}}^{r\dagger}e^{i(p+q)\cdot y}\right\} ,\label{eq:SC}
\end{eqnarray}
where auxiliary functions $\Phi_{(2)}$ are 
\begin{eqnarray}
\Phi_{(2)+-}^{sr;\mu\lambda\nu}(p,q) & = & \Phi_{(2)-+}^{sr;\mu\lambda\nu*}(p,q)\nonumber \\
 & = & \frac{1}{\sqrt{2E_{\mathbf{p}}}}\frac{1}{\sqrt{2E_{\mathbf{q}}}}i[p^{\mu}\epsilon_{s}^{\lambda*}({\bf p})-p^{\lambda}\epsilon_{s}^{\mu*}({\bf p})]\epsilon_{r}^{\nu}(\mathbf{q}),\label{eq:Phi_2pm}\\
\Phi_{(2)--}^{sr;\mu\lambda\nu}(p,q) & = & \Phi_{(2)++}^{sr;\mu\lambda\nu*}(p,q)\nonumber \\
 & = & \frac{1}{\sqrt{2E_{\mathbf{p}}}}\frac{1}{\sqrt{2E_{\mathbf{q}}}}i[p^{\mu}\epsilon_{s}^{\lambda}({\bf p})-p^{\lambda}\epsilon_{s}^{\mu}({\bf p})]\epsilon_{r}^{\nu}(\mathbf{q}).\label{eq:Phi_2mm}
\end{eqnarray}

Now we integrate over $y$ and $q$ in Eqs.~(\ref{eq:oam}) - (\ref{eq:SC})
and replace $p$ with $q$ as the integral variable, we obtain 
\begin{eqnarray}
\hat{L}^{\lambda\nu} & = & -x^{\lambda}\hat{P}^{\nu}+\frac{1}{2}n_{\mu}\int\frac{d^{3}q}{(2\pi)^{3}}\sum_{s,r}\nonumber \\
 &  & \times\left\{ \left[t_{y}\frac{q^{\lambda}}{E_{q}}\Phi_{(1)+-}^{sr;\mu\nu}(q,q)+ig_{i}^{\lambda}\partial_{q(1)}^{i}\Phi_{(1)+-}^{sr;\mu\nu}(q,q)\right]a_{{\bf q}}^{s\dagger}a_{{\bf q}}^{r}\right.\nonumber \\
 &  & +ig_{i}^{\lambda}\Phi_{(1)+-}^{sr;\mu\nu}(q,q)(\partial_{q}^{i}a_{{\bf q}}^{s\dagger})a_{{\bf q}}^{r}\nonumber \\
 &  & +\left[t_{y}\frac{q^{\lambda}}{E_{q}}\Phi_{(1)-+}^{sr;\mu\nu}(q,q)-ig_{i}^{\lambda}\partial_{q(1)}^{i}\Phi_{(1)-+}^{sr;\mu\nu}(q,q)\right]a_{{\bf q}}^{s}a_{{\bf q}}^{r\dagger}\nonumber \\
 &  & -ig_{i}^{\lambda}\Phi_{(1)-+}^{sr;\mu\nu}(q,q)(\partial_{q}^{i}a_{{\bf q}}^{s})a_{{\bf q}}^{r\dagger}\nonumber \\
 &  & +\left[t_{y}\frac{q^{\lambda}}{E_{q}}\Phi_{(1)--}^{sr;\mu\nu}(q,\tilde{q})-ig_{i}^{\lambda}\partial_{q(1)}^{i}\Phi_{(1)--}^{sr;\mu\nu}(q,\tilde{q})\right]e^{-2iE_{\mathbf{q}}t_{y}}a_{{\bf q}}^{s}a_{-{\bf q}}^{r}\nonumber \\
 &  & -ig_{i}^{\lambda}\Phi_{(1)--}^{sr;\mu\nu}(q,\tilde{q})e^{-2iE_{\mathbf{q}}t_{y}}(\partial_{q}^{i}a_{{\bf q}}^{s})a_{-{\bf q}}^{r}\nonumber \\
 &  & +\left[t_{y}\frac{q^{\lambda}}{E_{q}}\Phi_{(1)++}^{sr;\mu\nu}(q,\tilde{q})+ig_{i}^{\lambda}\partial_{q(1)}^{i}\Phi_{(1)++}^{sr;\mu\nu}(q,\tilde{q})\right]e^{2iE_{\mathbf{q}}t_{y}}a_{{\bf q}}^{s\dagger}a_{{\bf -q}}^{r\dagger}\nonumber \\
 &  & \left.+ig_{i}^{\lambda}\Phi_{(1)++}^{sr;\mu\nu}(q,\tilde{q})e^{2iE_{\mathbf{q}}t_{y}}(\partial_{q}^{i}a_{{\bf q}}^{s\dagger})a_{{\bf -q}}^{r\dagger}\right\} ,\label{eq:L_integrate_q}
\end{eqnarray}

\begin{eqnarray}
\hat{S}_{B}^{\lambda\nu} & = & -2n_{\mu}\int\frac{d^{3}q}{(2\pi)^{3}}\sum_{s,r}\left[\Phi_{(2)+-}^{sr;\mu\lambda\nu}(q,q)a_{\mathbf{q}}^{s\dagger}a_{\mathbf{q}}^{r}+\Phi_{(2)-+}^{sr;\mu\lambda\nu}(q,q)a_{{\bf q}}^{s}a_{{\bf q}}^{r\dagger}\right.\nonumber \\
 &  & \left.-\Phi_{(2)--}^{sr;\mu\lambda\nu}(q,\tilde{q})a_{{\bf q}}^{s}a_{-{\bf q}}^{r}e^{-2iE_{{\bf q}}t_{y}}-\Phi_{(2)++}^{sr;\mu\lambda\nu}(q,\tilde{q})a_{{\bf q}}^{s\dagger}a_{-{\bf q}}^{r\dagger}e^{2iE_{{\bf q}}t_{y}}\right],\label{eq:S1_integrate_q}
\end{eqnarray}
\begin{eqnarray}
\hat{S}_{C}^{\lambda\nu} & = & -n_{\mu}\int\frac{d^{3}q}{(2\pi)^{3}}\sum_{s,r}\left\{ \left[\Phi_{(2)+-}^{sr;\mu\lambda\nu}(q,q)-\Phi_{(2)+-}^{sr;\mu\nu\lambda}(q,q)\right]a_{\mathbf{q}}^{s\dagger}a_{\mathbf{q}}^{r}\right.\nonumber \\
 &  & +\left[\Phi_{(2)-+}^{sr;\mu\lambda\nu}(q,q)-\Phi_{(2)-+}^{sr;\mu\nu\lambda}(q,q)\right]a_{{\bf q}}^{s}a_{{\bf q}}^{r\dagger}\nonumber \\
 &  & -\left[\Phi_{(2)--}^{sr;\mu\lambda\nu}(q,\tilde{q})-\Phi_{(2)--}^{sr;\mu\nu\lambda}(q,\tilde{q})\right]a_{{\bf q}}^{s}a_{-{\bf q}}^{r}e^{-2iE_{{\bf q}}t_{y}}\nonumber \\
 &  & \left.-\left[\Phi_{(2)++}^{sr;\mu\lambda\nu}(q,\tilde{q})-\Phi_{(2)++}^{sr;\mu\nu\lambda}(q,\tilde{q})\right]a_{{\bf q}}^{s\dagger}a_{-{\bf q}}^{r\dagger}e^{2iE_{\mathbf{q}}t_{y}}\right\} ,\label{eq:S2_integrate_q-1}
\end{eqnarray}
where $\partial_{(1)}$ acts only on the first variable of $\Phi(1,2)$
in Eq. (\ref{eq:L_integrate_q}). We can also define $\partial_{(2)}$
which acts only on the second variable of $\Phi(1,2)$. The commutators
between OAM and spin tensor operators are 
\begin{eqnarray}
\left[\hat{L}^{\lambda\nu},a_{\mathbf{p}}^{s\dagger}\right] & = & -x^{\lambda}p^{\nu}a_{{\bf p}}^{s\dagger}\nonumber \\
 &  & +\frac{1}{2}\left\{ \left[in_{\mu}g_{i}^{\lambda}\left(\partial_{p(1)}^{i}\Phi_{(1)+-}^{rs;\mu\nu}(p,p)+\partial_{p(2)}^{i}\Phi_{(1)-+}^{sr;\mu\nu}(p,p)\right)\right.\right.\nonumber \\
 &  & \left.+2t_{y}\frac{1}{E_{p}}p^{\lambda}p^{\nu}\delta_{sr}+2i\delta_{sr}p^{\nu}g_{i}^{\lambda}\partial_{p}^{i}\right]a_{{\bf p}}^{r\dagger}\nonumber \\
 &  & \left.+in_{\mu}g_{i}^{\lambda}\left[\partial_{p(1)}^{i}\Phi_{(1)--}^{rs;\mu\nu}(\tilde{p},p)+\partial_{p(2)}^{i}\Phi_{(1)--}^{sr;\mu\nu}(p,\tilde{p})\right]e^{-2iE_{\mathbf{p}}t_{y}}a_{-{\bf p}}^{r}\right\} ,
\end{eqnarray}
\begin{eqnarray}
\left[\hat{L}^{\lambda\nu},a_{-\mathbf{p}}^{s}\right] & = & x^{\lambda}\tilde{p}^{\nu}a_{-{\bf p}}^{s}\nonumber \\
 &  & +\frac{1}{2}\left\{ \left[-in_{\mu}g_{i}^{\lambda}\left(\partial_{p(1)}^{i}\Phi_{(1)-+}^{rs;\mu\nu}(\tilde{p},\tilde{p})+\partial_{p(2)}^{i}\Phi_{(1)+-}^{sr;\mu\nu}(\tilde{p},\tilde{p})\right)\right.\right.\nonumber \\
 &  & \left.-2t_{y}\frac{\tilde{p}^{\lambda}\tilde{p}^{\nu}}{E_{p}}\delta_{sr}-2i\delta_{sr}\tilde{p}^{\nu}g_{i}^{\lambda}\partial_{p}^{i}\right]a_{-{\bf p}}^{r}\nonumber \\
 &  & \left.-in_{\mu}g_{i}^{\lambda}\left[\partial_{p(1)}^{i}\Phi_{(1)++}^{rs;\mu\nu}(p,\tilde{p})+\partial_{p(2)}^{i}\Phi_{(1)++}^{sr;\mu\nu}(\tilde{p},p)\right]e^{2iE_{p}t_{y}}a_{{\bf p}}^{r\dagger}\right\} ,
\end{eqnarray}
\begin{eqnarray}
\left[\hat{S}_{B}^{\lambda\nu},a_{\mathbf{p}}^{s\dagger}\right] & = & -2n_{\mu}\left\{ \left[\Phi_{(2)+-}^{rs;\mu\lambda\nu}(p,p)+\Phi_{(2)-+}^{sr;\mu\lambda\nu}(p,p)\right]a_{{\bf p}}^{r\dagger}\right.\nonumber \\
 &  & \left.-\left[\Phi_{(2)--}^{rs;\mu\lambda\nu}(\tilde{p},p)+\Phi_{(2)--}^{sr;\mu\lambda\nu}(p,\tilde{p})\right]e^{-2iE_{p}t_{y}}a_{-{\bf p}}^{r}\right\} ,\\
\left[\hat{S}_{B}^{\lambda\nu},a_{-\mathbf{p}}^{s}\right] & = & -2n_{\mu}\left\{ -\left[\Phi_{(2)+-}^{sr;\mu\lambda\nu}(\tilde{p},\tilde{p})+\Phi_{(2)-+}^{rs;\mu\lambda\nu}(\tilde{p},\tilde{p})\right]a_{-{\bf p}}^{r}\right.\nonumber \\
 &  & \left.+\left[\Phi_{(2)++}^{rs;\mu\lambda\nu}(p,\tilde{p})+\Phi_{(2)++}^{sr;\mu\lambda\nu}(\tilde{p},p)\right]e^{2iE_{p}t_{y}}a_{{\bf p}}^{r\dagger}\right\} ,\\
\nonumber 
\end{eqnarray}
\begin{eqnarray}
\left[\hat{S}_{C}^{\lambda\nu},a_{\mathbf{p}}^{s\dagger}\right] & = & -n_{\mu}\left\{ \left[\Phi_{(2)+-}^{rs;\mu\lambda\nu}(p,p)-\Phi_{(2)+-}^{rs;\mu\nu\lambda}(p,p)\right]a_{\mathbf{p}}^{r\dagger}\right.\nonumber \\
 &  & +\left[\Phi_{(2)-+}^{sr;\mu\lambda\nu}(p,p)-\Phi_{(2)-+}^{sr;\mu\nu\lambda}(p,p)\right]a_{{\bf p}}^{r\dagger}\nonumber \\
 &  & -\left[\Phi_{(2)--}^{rs;\mu\lambda\nu}(\tilde{p},p)-\Phi_{(2)--}^{rs;\mu\nu\lambda}(\tilde{p},p)\right]e^{-2iE_{p}t_{y}}a_{-{\bf p}}^{r}\nonumber \\
 &  & \left.-\left[\Phi_{(2)--}^{sr;\mu\lambda\nu}(p,\tilde{p})-\Phi_{(2)--}^{sr;\mu\nu\lambda}(p,\tilde{p})\right]e^{-2iE_{p}t_{y}}a_{-{\bf p}}^{r}\right\} ,\\
\left[\hat{S}_{C}^{\lambda\nu},a_{-\mathbf{p}}^{s}\right] & = & -n_{\mu}\left\{ -\left[\Phi_{(2)+-}^{sr;\mu\lambda\nu}(\tilde{p},\tilde{p})-\Phi_{(2)+-}^{sr;\mu\nu\lambda}(\tilde{p},\tilde{p})\right]a_{-\mathbf{p}}^{r}\right.\nonumber \\
 &  & -\left[\Phi_{(2)-+}^{rs;\mu\lambda\nu}(\tilde{p},\tilde{p})-\Phi_{(2)-+}^{rs;\mu\nu\lambda}(\tilde{p},\tilde{p})\right]a_{-{\bf p}}^{r}\nonumber \\
 &  & +\left[\Phi_{(2)++}^{sr;\mu\lambda\nu}(\tilde{p},p)-\Phi_{(2)++}^{sr;\mu\nu\lambda}(\tilde{p},p)\right]e^{2iE_{p}t_{y}}a_{{\bf p}}^{r\dagger}\nonumber \\
 &  & \left.+\left[\Phi_{(2)++}^{rs;\mu\lambda\nu}(p,\tilde{p})-\Phi_{(2)++}^{rs;\mu\nu\lambda}(p,\tilde{p})\right]e^{2iE_{p}t_{y}}a_{{\bf p}}^{r\dagger}\right\} 
\end{eqnarray}

Using Eqs.~(\ref{eq:B1_original},\ref{eq:B2_original}), the commutation
coefficients of Eqs.~(\ref{eq:commutation_B+},\ref{eq:commutation_B-})
become,
\begin{eqnarray}
\Xi_{(B)++}^{sr}(p) & = & -\Xi_{(B)--}^{sr*}(p\leftrightarrow\tilde{p})\nonumber \\
 & = & x^{\lambda}p^{\nu}(\partial_{\lambda}\beta_{\nu})\delta_{sr}\nonumber \\
 &  & -(\partial_{\lambda}\beta_{\nu})\left[\frac{1}{2}in_{\mu}g_{i}^{\lambda}\partial_{p}^{i}\left(\partial_{p(1)}^{i}\Phi_{(1)+-}^{rs;\mu\nu}(p,p)+\partial_{p(2)}^{i}\Phi_{(1)-+}^{sr;\mu\nu}(p,p)\right)\right.\nonumber \\
 &  & \left.+t_{y}\frac{1}{E_{p}}p^{\lambda}p^{\nu}\delta_{sr}+i\delta_{sr}p^{\nu}g_{i}^{\lambda}\partial_{p}^{i}-n_{\mu}\Phi_{(2)+-}^{rs;\mu\lambda\nu}(p,p)-n_{\mu}\Phi_{(2)-+}^{sr;\mu\lambda\nu}(p,p)\right],\label{eq:Xi_Bpp}
\end{eqnarray}
\begin{eqnarray}
\Xi_{(B)+-}^{sr}(p) & = & -\Xi_{(B)-+}^{sr*}(p\leftrightarrow\tilde{p})\nonumber \\
 & = & -(\partial_{\lambda}\beta_{\nu})e^{-2iE_{\mathbf{p}}t_{y}}n_{\mu}\left\{ \frac{1}{2}ig_{i}^{\lambda}\left[\partial_{p(1)}^{i}\Phi_{(1)--}^{rs;\mu\nu}(\tilde{p},p)+\partial_{p(2)}^{i}\Phi_{(1)--}^{sr;\mu\nu}(p,\tilde{p})\right]\right.\nonumber \\
 &  & \left.+\Phi_{(2)--}^{rs;\mu\lambda\nu}(\tilde{p},p)+\Phi_{(2)--}^{sr;\mu\lambda\nu}(p,\tilde{p})\right\} ,\label{eq:Xi_Bpm}
\end{eqnarray}
\begin{eqnarray}
\Xi_{(C)++}^{sr}(p) & = & -\Xi_{(C)--}^{sr*}(p\leftrightarrow\tilde{p})\nonumber \\
 & = & x^{\lambda}p^{\nu}(\partial_{\lambda}\beta_{\nu})\delta_{sr}\nonumber \\
 &  & -(\partial_{\lambda}\beta_{\nu})\left\{ \frac{1}{2}in_{\mu}g_{i}^{\lambda}\partial_{p}^{i}\left[\partial_{p(1)}^{i}\Phi_{(1)+-}^{rs;\mu\nu}(p,p)+\partial_{p(2)}^{i}\Phi_{(1)-+}^{sr;\mu\nu}(p,p)\right]\right.\nonumber \\
 &  & \left.+t_{y}\frac{1}{E_{p}}p^{\lambda}p^{\nu}\delta_{sr}+i\delta_{sr}p^{\nu}g_{i}^{\lambda}\partial_{p}^{i}\right\} \nonumber \\
 &  & -\frac{1}{2}\Omega_{\lambda\nu}n_{\mu}\left[\Phi_{(2)+-}^{rs;\mu\lambda\nu}(p,p)-\Phi_{(2)+-}^{rs;\mu\nu\lambda}(p,p)+\Phi_{(2)-+}^{sr;\mu\lambda\nu}(p,p)-\Phi_{(2)-+}^{sr;\mu\nu\lambda}(p,p)\right],\label{eq:Xi_Cpp}
\end{eqnarray}
\begin{eqnarray}
\Xi_{(C)+-}^{sr}(p) & = & -\Xi_{(C)-+}^{sr*}(p\leftrightarrow\tilde{p})\nonumber \\
 & = & -i\frac{1}{2}(\partial_{\lambda}\beta_{\nu})e^{-2iE_{\mathbf{p}}t_{y}}n_{\mu}g_{i}^{\lambda}\left[\partial_{p(1)}^{i}\Phi_{(1)--}^{rs;\mu\nu}(\tilde{p},p)+\partial_{p(2)}^{i}\Phi_{(1)--}^{sr;\mu\nu}(p,\tilde{p})\right]\nonumber \\
 &  & +\frac{1}{2}\Omega_{\lambda\nu}e^{-2iE_{\mathbf{p}}t_{y}}n_{\mu}\nonumber \\
 &  & \times\left[\Phi_{(2)--}^{rs;\mu\lambda\nu}(\tilde{p},p)-\Phi_{(2)--}^{rs;\mu\nu\lambda}(\tilde{p},p)+\Phi_{(2)--}^{sr;\mu\lambda\nu}(p,\tilde{p})-\Phi_{(2)--}^{sr;\mu\nu\lambda}(p,\tilde{p})\right].\label{eq:Xi_Cpm}
\end{eqnarray}

We use Eqs.~(\ref{eq:commutation_A+})-(\ref{eq:commutation_B-})
and (\ref{eq:Baker-Hausdorff_formular}) to evaluate Eqs.~(\ref{eq:rho_a^dagger_rho})
and (\ref{eq:rho_a_rho}), the coefficients are obtained as 
\begin{eqnarray}
D_{++}^{sr} & = & -e^{\beta\cdot p}\int_{0}^{1}d\lambda e^{-\lambda\beta\cdot p}\Xi_{++}^{sr}e^{\lambda\beta\cdot p}\nonumber \\
 &  & +\sum_{s^{\prime}}e^{\beta\cdot p}\int_{0}^{1}d\lambda_{1}\int_{0}^{\lambda_{1}}d\lambda_{2}e^{-\lambda_{1}\beta\cdot p}\Xi_{++}^{ss^{\prime}}e^{\lambda_{1}\beta\cdot p}e^{-\lambda_{2}\beta\cdot p}\Xi_{++}^{s^{\prime}r}e^{\lambda_{2}\beta\cdot p}\nonumber \\
 &  & +\sum_{s^{\prime}}e^{\beta\cdot p}\int_{0}^{1}d\lambda_{1}\int_{0}^{\lambda_{1}}d\lambda_{2}e^{-\lambda_{1}\beta\cdot p}\Xi_{+-}^{ss^{\prime}}e^{-\lambda_{1}\beta\cdot\tilde{p}}e^{\lambda_{2}\beta\cdot\tilde{p}}\Xi_{-+}^{s^{\prime}r}e^{\lambda_{2}\beta\cdot p},\label{eq:Dpp}\\
D_{+-}^{sr} & = & -e^{\beta\cdot p}\int_{0}^{1}d\lambda e^{-\lambda\beta\cdot p}\Xi_{+-}^{sr}e^{-\lambda\beta\cdot\tilde{p}}\nonumber \\
 &  & +\sum_{s^{\prime}}e^{\beta\cdot p}\int_{0}^{1}d\lambda_{1}\int_{0}^{\lambda_{1}}d\lambda_{2}e^{-\lambda_{1}\beta\cdot p}\Xi_{++}^{ss^{\prime}}e^{\lambda_{1}\beta\cdot p}e^{-\lambda_{2}\beta\cdot p}\Xi_{+-}^{s^{\prime}r}e^{-\lambda_{2}\beta\cdot\tilde{p}}\nonumber \\
 &  & +\sum_{s^{\prime}}e^{\beta\cdot p}\int_{0}^{1}d\lambda_{1}\int_{0}^{\lambda_{1}}d\lambda_{2}e^{-\lambda_{1}\beta\cdot p}\Xi_{+-}^{ss^{\prime}}e^{-\lambda_{1}\beta\cdot\tilde{p}}e^{\lambda_{2}\beta\cdot\tilde{p}}\Xi_{--}^{s^{\prime}r}e^{-\lambda_{2}\beta\cdot\tilde{p}},\label{eq:Dpm}\\
D_{--}^{sr} & = & -e^{-\beta\cdot\tilde{p}}\int_{0}^{1}d\lambda e^{\lambda\beta\cdot\tilde{p}}\Xi_{--}^{sr}e^{-\lambda\beta\cdot\tilde{p}}\nonumber \\
 &  & +\sum_{s^{\prime}}e^{-\beta\cdot\tilde{p}}\int_{0}^{1}d\lambda_{1}\int_{0}^{\lambda_{1}}d\lambda_{2}e^{\lambda_{1}\beta\cdot\tilde{p}}\Xi_{--}^{ss^{\prime}}e^{-\lambda_{1}\beta\cdot\tilde{p}}e^{\lambda_{2}\beta\cdot\tilde{p}}\Xi_{--}^{s^{\prime}r}e^{-\lambda_{2}\beta\cdot\tilde{p}}\nonumber \\
 &  & +\sum_{s^{\prime}}e^{-\beta\cdot\tilde{p}}\int_{0}^{1}d\lambda_{1}\int_{0}^{\lambda_{1}}d\lambda_{2}e^{\lambda_{1}\beta\cdot\tilde{p}}\Xi_{-+}^{ss^{\prime}}e^{\lambda_{1}\beta\cdot p}e^{-\lambda_{2}\beta\cdot p}\Xi_{+-}^{s^{\prime}r}e^{-\lambda_{2}\beta\cdot\tilde{p}},\label{eq:Dmm}\\
D_{-+}^{sr} & = & -e^{-\beta\cdot\tilde{p}}\int_{0}^{1}d\lambda e^{\lambda\beta\cdot\tilde{p}}\Xi_{-+}^{sr}e^{\lambda\beta\cdot p}\nonumber \\
 &  & +\sum_{s^{\prime}}e^{-\beta\cdot\tilde{p}}\int_{0}^{1}d\lambda_{1}\int_{0}^{\lambda_{1}}d\lambda_{2}e^{\lambda_{1}\beta\cdot\tilde{p}}\Xi_{--}^{ss^{\prime}}e^{-\lambda_{1}\beta\cdot\tilde{p}}e^{\lambda_{2}\beta\cdot\tilde{p}}\Xi_{-+}^{s^{\prime}r}e^{\lambda_{2}\beta\cdot p}\nonumber \\
 &  & +\sum_{s^{\prime}}e^{-\beta\cdot\tilde{p}}\int_{0}^{1}d\lambda_{1}\int_{0}^{\lambda_{1}}d\lambda_{2}e^{\lambda_{1}\beta\cdot\tilde{p}}\Xi_{-+}^{ss^{\prime}}e^{\lambda_{1}\beta\cdot p}e^{-\lambda_{2}\beta\cdot p}\Xi_{++}^{s^{\prime}r}e^{\lambda_{2}\beta\cdot p}.\label{eq:Dmp}
\end{eqnarray}


\section{Coefficients for Eqs.~(\ref{eq:Sigma_C_01}, \ref{eq:Sigma_B_01})
\protect\label{sec:Coefficients}}

Here, we list the coefficients in Eqs.~(\ref{eq:Sigma_B_01}) and
(\ref{eq:Sigma_C_01}):
\begin{eqnarray}
(\Xi_{(B)\beta\beta}^{ss})_{\lambda\nu,\rho\sigma} & = & -\tilde{\Delta}_{\lambda\nu}\tilde{\Delta}_{\rho\sigma}-\tilde{\Delta}_{\lambda\nu}(\epsilon_{\rho}^{s*}\epsilon_{\sigma}^{s}+\epsilon_{\sigma}^{s*}\epsilon_{\rho}^{s})-\tilde{\Delta}_{\rho\sigma}(\epsilon_{\lambda}^{s*}\epsilon_{\nu}^{s}+\epsilon_{\lambda}^{s}\epsilon_{\nu}^{s*})\nonumber \\
 &  & +(\epsilon_{\rho}^{s*}\epsilon_{\nu}^{s}\tilde{\Delta}_{\lambda\sigma}+\epsilon_{\sigma}^{s*}\epsilon_{\nu}^{s}\tilde{\Delta}_{\lambda\rho}+\epsilon_{\rho}^{s*}\epsilon_{\lambda}^{s}\tilde{\Delta}_{\nu\sigma}+\epsilon_{\sigma}^{s*}\epsilon_{\lambda}^{s}\tilde{\Delta}_{\nu\rho})\nonumber \\
 &  & -\frac{m}{E_{\mathbf{k}}(E_{\mathbf{k}}+m)}\tilde{\Delta}_{\lambda\nu}\left[\frac{k\cdot\epsilon_{s}^{*}}{m}(\epsilon_{\rho}^{s}\bar{k}_{\sigma}+\epsilon_{\sigma}^{s}\bar{k}_{\rho})+\frac{k\cdot\epsilon_{s}}{m}(\epsilon_{\rho}^{s*}\bar{k}_{\sigma}+\epsilon_{\sigma}^{s*}\bar{k}_{\rho})\right]\nonumber \\
 &  & -\frac{m}{E_{\mathbf{k}}(E_{\mathbf{k}}+m)}\tilde{\Delta}_{\rho\sigma}\left[\frac{k\cdot\epsilon_{s}^{*}}{m}(\epsilon_{\lambda}^{s}\bar{k}_{\nu}+\epsilon_{\nu}^{s}\bar{k}_{\lambda})+\frac{k\cdot\epsilon_{s}}{m}(\epsilon_{\lambda}^{s*}\bar{k}_{\nu}+\epsilon_{\nu}^{s*}\bar{k}_{\lambda})\right]\nonumber \\
 &  & +\frac{m}{E_{\mathbf{k}}(E_{\mathbf{k}}+m)}\frac{k\cdot\epsilon_{s}^{*}}{m}(\tilde{\Delta}_{\lambda\rho}\bar{k}_{\sigma}\epsilon_{\nu}^{s}+\tilde{\Delta}_{\nu\rho}\bar{k}_{\sigma}\epsilon_{\lambda}^{s}+\tilde{\Delta}_{\lambda\sigma}\bar{k}_{\rho}\epsilon_{\nu}^{s}+\tilde{\Delta}_{\nu\sigma}\bar{k}_{\rho}\epsilon_{\lambda}^{s})\nonumber \\
 &  & +\frac{m}{E_{\mathbf{k}}(E_{\mathbf{k}}+m)}\frac{k\cdot\epsilon_{s}}{m}(\tilde{\Delta}_{\rho\lambda}\bar{k}_{\nu}\epsilon_{\sigma}^{s*}+\tilde{\Delta}_{\lambda\sigma}\bar{k}_{\nu}\epsilon_{\rho}^{s*}+\tilde{\Delta}_{\nu\sigma}\bar{k}_{\lambda}\epsilon_{\rho}^{s*}+\tilde{\Delta}_{\rho\nu}\bar{k}_{\lambda}\epsilon_{\sigma}^{s*})\nonumber \\
 &  & +\frac{m^{2}}{E_{\mathbf{k}}^{2}(E_{\mathbf{k}}+m)^{2}}\frac{k\cdot\epsilon_{s}^{*}}{m}\frac{k\cdot\epsilon_{s}}{m}(\tilde{\Delta}_{\lambda\rho}\bar{k}_{\nu}\bar{k}_{\sigma}+\tilde{\Delta}_{\lambda\sigma}\bar{k}_{\nu}\bar{k}_{\rho}+\tilde{\Delta}_{\nu\rho}\bar{k}_{\lambda}\bar{k}_{\sigma}\nonumber \\
 &  & +\tilde{\Delta}_{\nu\sigma}\bar{k}_{\lambda}\bar{k}_{\rho}-2\tilde{\Delta}_{\lambda\nu}\bar{k}_{\rho}\bar{k}_{\sigma}-2\tilde{\Delta}_{\rho\sigma}\bar{k}_{\lambda}\bar{k}_{\nu}),\\
(\Xi_{(C)\beta\beta}^{ss})_{\lambda\nu,\rho\sigma} & = & -\tilde{\Delta}_{\lambda\nu}\tilde{\Delta}_{\rho\sigma}\delta_{ss}+\frac{\tilde{k}_{\nu}\tilde{k}_{\sigma}}{E_{\mathbf{k}}^{2}}\frac{\bar{k}^{2}}{m^{2}}\epsilon_{\lambda}^{s}\epsilon_{\rho}^{s*}\nonumber \\
 &  & -\frac{k_{\sigma}}{E_{\mathbf{k}}}\frac{k\cdot\epsilon_{s}^{*}}{m}\epsilon_{\rho}^{s}\tilde{\Delta}_{\lambda\nu}+\frac{\tilde{k}_{\sigma}}{E_{\mathbf{k}}}\frac{k\cdot\epsilon_{s}}{m}\epsilon_{\rho}^{s*}\left[\left(\Delta_{\lambda\nu}-\frac{\bar{k}_{\lambda}\bar{k}_{\nu}}{E_{\mathbf{k}}^{2}}\right)-\frac{\bar{k}_{\lambda}\tilde{k}_{\nu}}{mE_{\mathbf{k}}}\right]\nonumber \\
 &  & -\frac{k_{\nu}}{E_{\mathbf{k}}}\frac{k\cdot\epsilon_{s}}{m}\epsilon_{\lambda}^{s*}\tilde{\Delta}_{\rho\sigma}+\frac{\tilde{k}_{\nu}}{E_{\mathbf{k}}}\frac{k\cdot\epsilon_{s}^{*}}{m}\epsilon_{\lambda}^{s}\left[\left(\Delta_{\rho\sigma}-\frac{\bar{k}_{\rho}\bar{k}_{\sigma}}{E_{\mathbf{k}}^{2}}\right)-\frac{\bar{k}_{\rho}\tilde{k}_{\sigma}}{mE_{\mathbf{k}}}\right]\nonumber \\
 &  & -(m\Delta_{\lambda\nu}-u_{\nu}\bar{k}_{\lambda})\frac{2\bar{k}_{\rho}\bar{k}_{\sigma}}{E_{\mathbf{k}}^{2}(E_{\mathbf{k}}+m)}\frac{k\cdot\epsilon_{s}^{*}}{m}\frac{k\cdot\epsilon_{s}}{m}\nonumber \\
 &  & -(m\Delta_{\rho\sigma}-u_{\sigma}\bar{k}_{\rho})\frac{2\bar{k}_{\lambda}\bar{k}_{\nu}}{E_{\mathbf{k}}^{2}(E_{\mathbf{k}}+m)}\frac{k\cdot\epsilon_{s}^{*}}{m}\frac{k\cdot\epsilon_{s}}{m}+\frac{k_{\nu}k_{\sigma}}{E_{\mathbf{k}}^{2}}\Delta_{\lambda\rho}\frac{k\cdot\epsilon_{s}^{*}}{m}\frac{k\cdot\epsilon_{s}}{m},\\
(\Xi_{(C)\beta\Omega}^{ss})_{\lambda\nu,\rho\sigma} & = & \frac{\tilde{k}_{\nu}\tilde{k}_{\sigma}}{E_{\mathbf{k}}^{2}}\frac{\bar{k}^{2}}{m^{2}}\epsilon_{\rho}^{s*}\epsilon_{\lambda}^{s}-\frac{\tilde{k}_{\nu}k_{\sigma}}{E_{\mathbf{k}}^{2}}\frac{\bar{k}_{\rho}}{m}\frac{k\cdot\epsilon_{s}^{*}}{m}\epsilon_{\lambda}^{s}-\tilde{\Delta}_{\lambda\nu}\frac{k_{\sigma}}{E_{\mathbf{k}}}\frac{k\cdot\epsilon_{s}^{*}}{m}\epsilon_{\rho}^{s}\nonumber \\
 &  & +\frac{\tilde{k}_{\sigma}}{E_{\mathbf{k}}}\frac{k\cdot\epsilon_{s}}{m}\epsilon_{\rho}^{s*}\left(\Delta_{\lambda\nu}-\frac{\bar{k}_{\lambda}\bar{k}_{\nu}}{E_{\mathbf{k}}^{2}}\right)-\frac{\bar{k}_{\lambda}}{m}\frac{\tilde{k}_{\nu}\tilde{k}_{\sigma}}{E_{\mathbf{k}}^{2}}\frac{k\cdot\epsilon_{s}}{m}\epsilon_{\rho}^{s*}\nonumber \\
 &  & +\Delta_{\lambda\rho}\frac{k_{\nu}k_{\sigma}}{E_{\mathbf{k}}^{2}}\frac{k\cdot\epsilon_{s}^{*}}{m}\frac{k\cdot\epsilon_{s}}{m}+\bar{k}_{\rho}k_{\sigma}\bar{k}_{\lambda}\bar{k}_{\nu}\frac{2}{E_{\mathbf{k}}^{3}(E_{\mathbf{k}}+m)}\frac{k\cdot\epsilon_{s}^{*}}{m}\frac{k\cdot\epsilon_{s}}{m},\\
(\Xi_{(C)\Omega\Omega}^{ss})_{\lambda\nu,\rho\sigma} & = & \frac{\tilde{k}_{\nu}\tilde{k}_{\sigma}}{E_{\mathbf{k}}^{2}}\frac{\bar{k}^{2}}{m^{2}}\epsilon_{\rho}^{s*}\epsilon_{\lambda}^{s}-\frac{\tilde{k}_{\nu}k_{\sigma}}{E_{\mathbf{k}}^{2}}\frac{\bar{k}_{\rho}}{m}\frac{k\cdot\epsilon_{s}^{*}}{m}\epsilon_{\lambda}^{s}\nonumber \\
 &  & -\frac{k_{\nu}\tilde{k}_{\sigma}}{E_{\mathbf{k}}^{2}}\frac{\bar{k}_{\lambda}}{m}\frac{k\cdot\epsilon_{s}}{m}\epsilon_{\rho}^{s*}+\frac{k_{\nu}k_{\sigma}}{E_{\mathbf{k}}^{2}}\frac{k\cdot\epsilon_{s}^{*}}{m}\frac{k\cdot\epsilon_{s}}{m}\Delta_{\lambda\rho},
\end{eqnarray}
where we defined 
\begin{eqnarray}
\tilde{\Delta}_{\mu\nu} & = & \Delta_{\mu\nu}+\frac{\bar{k}_{\mu}\bar{k}_{\nu}}{E_{\mathbf{k}}^{2}}.
\end{eqnarray}


\bibliographystyle{h-physrev}
\bibliography{qkt-ref}

\end{document}